\documentclass[longauth]{aa}  

\usepackage[version=4]{mhchem}
\usepackage{natbib}
\bibpunct{(}{)}{;}{a}{}{,} 

\usepackage{graphicx}
\usepackage{textcomp}
\usepackage[hidelinks]{hyperref}
\usepackage{subfig}
\usepackage{fancyhdr}
\hypersetup{colorlinks, citecolor=blue, linkcolor=blue, urlcolor=blue}

\usepackage{txfonts}
\begin{document}

   \title{Large Interferometer For Exoplanets (LIFE). XIV. \\Finding terrestrial protoplanets in the galactic neighborhood}
   \titlerunning{LIFE XIV. Finding terrestrial protoplanets in the galactic neighborhood}
    \author{    
                Lorenzo Cesario~\,\orcid{0009-0009-7228-7809}\inst{\ref{inst1}}
                \and Tim Lichtenberg~\,\orcid{0000-0002-3286-7683}\inst{\ref{inst1}}
                \and Eleonora Alei~\,\orcid{0000-0002-0006-1175} \inst{\ref{inst2}}
                \and Óscar Carrión-González~\,\orcid{0000-0001-8466-9622}\inst{\ref{inst3}}
                \and Felix A. Dannert~\,\orcid{0000-0002-5476-2663}\inst{\ref{inst9},\ref{inst4}}
                \and Denis Defrère~\,\orcid{0000-0003-3499-2506}\inst{\ref{inst5}}  
                \and Steve Ertel~\,\orcid{0000-0002-2314-7289}\inst{\ref{inst6a},\ref{inst6b}}  
                \and Andrea Fortier~\,\orcid{0000-0001-8450-3374}\inst{\ref{inst7}}
                \and A. García Muñoz~\,\orcid{0000-0003-1756-4825}\inst{\ref{inst8}}
                \and Adrian M. Glauser~\,\orcid{0000-0001-9250-1547}\inst{\ref{inst9}}
                \and Jonah T. Hansen~\,\orcid{0000-0003-3992-342X}\inst{\ref{inst9},\ref{inst4}}
                \and Ravit Helled~\,\orcid{0000-0001-5555-2652}\inst{\ref{inst11}}
                \and Philipp A. Huber~\,\orcid{0009-0000-7417-3201}\inst{\ref{inst9}}
                \and Michael J. Ireland~\,\orcid{0000-0002-6194-043X}\inst{\ref{inst12}}
                \and Jens Kammerer~\,\orcid{0000-0003-2769-0438} \inst{\ref{inst13}}
                \and Romain Laugier~\,\orcid{0000-0002-2215-9413}\inst{\ref{inst5}}
                \and Jorge Lillo-Box~\,\orcid{0000-0003-3742-1987}\inst{\ref{inst14}}
                \and Franziska Menti~\,\orcid{0009-0006-0336-6827}\inst{\ref{inst9}}
                \and Michael R. Meyer~\,\orcid{0000-0003-1227-3084}\inst{\ref{inst15}}
                \and Lena Noack~\,\orcid{0000-0001-8817-1653}\inst{\ref{inst16}}
                \and Sascha P. Quanz~\,\orcid{0000-0003-3829-7412} \inst{\ref{inst9},\ref{inst17},\ref{inst4}}
                \and Andreas Quirrenbach~\,\orcid{0000-0002-3302-1962}\inst{\ref{inst19}}
                \and Sarah Rugheimer~\,\orcid{0000-0003-1620-7658}\inst{\ref{inst20}}
                \and Floris van der Tak~\,\orcid{0000-0002-8942-1594}\inst{\ref{inst21}}
                \and Haiyang S. Wang~\,\orcid{0000-0001-8618-3343}\inst{\ref{inst9}, \ref{inst22},\ref{inst23}}
                \and Marius Anger~\,\orcid{0000-0001-7037-425X}\inst{\ref{inst24}}
                \and Olga Balsalobre-Ruza~\,\orcid{0000-0002-6712-3035}\inst{\ref{inst14}}
                \and Surendra Bhattarai~\,\orcid{0000-0001-9026-8622}\inst{\ref{inst25}}
                \and Marrick Braam~\,\orcid{0000-0002-9076-2361}\inst{\ref{inst5},\ref{inst26},\ref{inst27}}
                \and Amadeo Castro-Gonz\'{a}lez~\,\orcid{0000-0001-7439-3618}\inst{\ref{inst14}}
                \and Charles S. Cockell~\,\orcid{0000-0003-3662-0503}\inst{\ref{inst28}}
                \and Tereza Constantinou~\,\orcid{0000-0002-2129-1340}\inst{\ref{inst29}}
                \and Gabriele Cugno~\,\orcid{0000-0001-7255-3251}\inst{\ref{inst30}}
                \and Jeanne Davoult~\,\orcid{0000-0002-6177-2085}\inst{\ref{inst31}}
                \and Manuel Güdel~\,\orcid{0000-0001-9818-0588}\inst{\ref{inst32}}
                \and Nina Hernitschek~\,\orcid{0000-0003-1681-0430}\inst{\ref{inst33}}
                \and Sasha Hinkley~\,\orcid{0000-0001-8074-2562}\inst{\ref{inst34}}
                \and Satoshi Itoh~\,\orcid{0000-0003-2690-7092}\inst{\ref{inst35}}
                \and Markus Janson~\,\orcid{0000-0001-8345-593X}\inst{\ref{inst36}}
                \and Anders Johansen~\,\orcid{0000-0002-5893-6165}\inst{\ref{inst37}}
                \and Hugh R. A. Jones~\,\orcid{0000-0003-0433-3665}\inst{\ref{inst38}}
                \and Stephen R. Kane~\,\orcid{0000-0002-7084-0529}\inst{\ref{inst39}}
                \and Tim A. van Kempen~\,\orcid{0000-0001-8178-5054}\inst{\ref{inst48}}
                \and Kristina G. Kislyakova~\,\orcid{0000-0003-4680-6774}\inst{\ref{inst32}}
                \and Judith Korth~\,\orcid{0000-0002-0076-6239}\inst{\ref{inst41}}
                \and Andjelka B. Kova{\v c}evi{\' c}~\,\orcid{0000-0001-5139-1978}\inst{\ref{inst42}}
                \and Stefan Kraus~\,\orcid{0000-0001-6017-8773}\inst{\ref{inst43}}
                \and Rolf Kuiper~\,\orcid{0000-0003-2309-8963}\inst{\ref{inst44}}
                \and Joice Mathew~\,\orcid{0000-0003-0459-5964}\inst{\ref{inst45}}
                \and Taro Matsuo~\,\orcid{0000-0001-7694-5885}\inst{\ref{inst46}}
                \and Yamila Miguel~\,\orcid{0000-0002-0747-8862}\inst{\ref{inst47}}
                \and Michiel Min~\,\orcid{0000-0001-5778-0376}\inst{\ref{inst48}}
                \and Ramon Navarro~\,\orcid{0009-0009-7018-2819}\inst{\ref{inst49}}
                \and Ramses M. Ramirez~\,\orcid{0000-0001-7553-8444}\inst{\ref{inst50}}
                \and Heike Rauer~\,\orcid{0000-0002-6510-1828}\inst{\ref{inst51}}
                \and Berke Vow Ricketti ~\,\orcid{0000-0001-9701-5660}\inst{\ref{inst58},\ref{inst59}}
                \and Amedeo Romagnolo~\,\orcid{0000-0001-9583-4339}\inst{\ref{inst52}}
                \and Martin Schlecker~\,\orcid{0000-0001-8355-2107}\inst{\ref{inst53}}
                \and Evan L. Sneed~\,\orcid{0000-0001-5290-1001}\inst{\ref{inst39}}
                \and Vito Squicciarini~\,\orcid{0000-0002-3122-6809}\inst{\ref{inst3},\ref{inst55}}
                \and Keivan G. Stassun~\,\orcid{0000-0002-3481-9052}\inst{\ref{inst56}}
                \and Motohide Tamura~\,\orcid{0000-0002-6510-0681}\inst{\ref{inst35}}
                \and Daniel Viudez-Moreiras~\,\orcid{0000-0001-8442-3788}\inst{\ref{inst14}}
                \and Robin D. Wordsworth~\,\orcid{0000-0003-1127-8334}\inst{\ref{inst57}}
                \and the LIFE Collaboration\inst{\ref{instfinal}}
            }

    \institute{
                Kapteyn Astronomical Institute, University of Groningen, P.O. Box 800, 9700 AV Groningen, The Netherlands\\\email{l.cesario@rug.nl; tim.lichtenberg@rug.nl}\label{inst1} 
                \and NPP Fellow, NASA Goddard Space Flight Center, Greenbelt, MD, USA \label{inst2}
                \and LESIA, Observatoire de Paris, Universit\'e PSL, CNRS, Sorbonne Universit\'e, Universit\'e Paris Cit\'e, 5 place Jules Janssen, 92195 Meudon, France \label{inst3}
                \and Institute for Particle Physics and Astrophysics, ETH Zurich, Wolfgang-Pauli-Str. 27, CH-8093 Zurich, Switzerland \label{inst9}
                \and National Center of Competence in Research PlanetS (www.nccr-planets.ch)\label{inst4}
                \and Institute of Astronomy, KU Leuven, Celestijnenlaan 200D, 3001 Leuven, Belgium \label{inst5}
                \and Department of Astronomy and Steward Observatory, The University of Arizona, 933 N Cherry Ave, Tucson, AZ 85721, USA\label{inst6a} 
                \and Large Binocular Telescope Observatory, The University of Arizona, 933 N Cherry Ave, Tucson, AZ 85721, USA \label{inst6b}
                \and Weltraumforschung und Planetologie, Physikalisches Institut, University of Bern, Gesellschaftsstrasse 6, 3012 Bern, Switzerland \\ Center for Space and Habitability, University of Bern, Gesellschaftsstrasse 6, 3012 Bern, Switzerland\label{inst7}
                \and Université Paris-Saclay, Université Paris Cité, CEA, CNRS, AIM, 91191, Gif-sur-Yvette, France\label{inst8}
                \and National Center of Competence in Research PlanetS, Switzerland (www.nccr-planets.ch) \label{inst10}
                \and Department of Astrophysics, University of Zurich, Winterthurerstr. 190, CH-8057 Zurich \label{inst11}
                \and Research School of Astronomy and Astrophysics, Australian National University, Canberra 2611, Australia\label{inst12}
                \and European Southern Observatory, Karl-Schwarzschild-Straße 2, 85748 Garching, Germany \label{inst13}
                \and Centro de Astrobiolog\'ia (CAB), CSIC-INTA, ESAC campus, Camino Bajo del Castillo s/n, 28692, Villanueva de la Ca\~nada (Madrid), Spain \label{inst14}
                \and Department of Astronomy, The University of Michigan, Ann Arbor, USA \label{inst15}
                \and Freie Universität Berlin, Institute of Geological Sciences, Berlin \label{inst16}
                \and ETH Zurich, Department of Earth and Planetary Sciences, Sonneggstrasse 5, 8092 Zurich, Switzerland \label{inst17}
                \and National Center of Competence in Research "PlanetS", Switzerland \label{inst18}
                \and Landessternwarte, Zentrum für Astronomie der Universität Heidelberg, Königstuhl 12, 69117 Heidelberg, Germany\label{inst19}
                \and Dept of Physics and Astronomy, York University, 4700 Keele St, Toronto, M3J 1P3, Canada \label{inst20}
                \and SRON Netherlands Institute for Space Research; Kapteyn Astronomical Institute, University of Groningen \label{inst21}
                \and Institute of Geochemistry and Petrology, ETH Zurich, Clausiusstrasse 25, 8092 Zurich, Switzerland \label{inst22}
                \and Center for Star and Planet Formation, Globe Institute, University of Copenhagen, Øster Voldgade 5-7, 1350 Copenhagen, Denmark \label{inst23}
                \and Department of Electronics and Nanoengineering, Aalto University \label{inst24}
                \and Department of Physical Sciences, Indian Institute of Science Education and Research Kolkata, Mohanpur 741246, West Bengal, India \label{inst25}
                \and School of GeoSciences, University of Edinburgh, Edinburgh, EH9 3FF, UK \label{inst26}
                \and Centre for Exoplanet Science, University of Edinburgh, Edinburgh, EH9 3FD, UK \label{inst27}
                \and UK Centre for Astrobiology, School of Physics and Astronomy, University of Edinburgh, United Kingdom \label{inst28}
                \and Institute of Astronomy, University of Cambridge, Cambridge, CB3 0HA, UK \label{inst29}
                \and Department of Astronomy, University of Michigan, Ann Arbor, MI 48109, USA \label{inst30}
                \and Space research \& Planetary Sciences (WP), Universität Bern, Gesellschaftsstrasse 6, 3012 Bern, Switzerland \label{inst31}
                \and Department of Astrophysics, University of Vienna, Türkenschanzstr. 17, 1180 Vienna, Austria \label{inst32}
                \and Universidad de Antofagasta \label{inst33}
                \and Department of Physics \& Astronomy, University of Exeter, United Kingdom \label{inst34}
                \and The University of Tokyo, Astrobiology Center \label{inst35}
                \and Department of Astronomy, Stockholm University, AlbaNova University Center, 10691 Stockholm, Sweden \label{inst36}
                \and Globe Institute, University of Copenhagen \label{inst37}
                \and Centre for Astrophysics Research, University of Hertfordshire, Hatfield, Hertfordshire, AL10 9AB, UK \label{inst38}
                \and Department of Earth and Planetary Sciences, University of California, Riverside, CA 92521, USA \label{inst39}
                \and Lund Observatory, Division of Astrophysics, Department of Physics, Lund University, Box 118, 22100 Lund, Sweden \label{inst41}
                \and University of Belgrade, Faculty of Mathematics, Department of astronomy, Studentski trg 16, Belgrade 11000, Serbia \label{inst42}
                \and Astrophysics Group, Department of Physics \& Astronomy, University of Exeter, Stocker Road, Exeter, EX4 4QL, UK \label{inst43}
                \and Fakultät für Physik, Universität Duisburg–Essen, Lotharstraße 1, 47057 Duisburg, Germany \label{inst44}
                \and Advanced Instrumentation and Technology Centre, Research School of Astronomy and Astrophysics, Australian National University, Canberra, ACT 2611, Australia \label{inst45}
                \and Nagoya University \label{inst46}
                \and SRON / Leiden Observatory \label{inst47}
                \and SRON Netherlands Institute for Space Research, Leiden, The Netherlands \label{inst48}
                \and NOVA Optical Infrared Instrumentation Group at ASTRON, Oude Hoogeveensedijk 4, NL-7991 PD Dwingeloo, the Netherlands \label{inst49}
                \and University of Central Florida \label{inst50}
                \and Institut fuer Planetenforschung, DLR, and FU Berlin, Germany \label{inst51}
                \and Disruptive Space Technology Centre, RAL Space, STFC-Rutherford Appleton Laboratory, Didcot, United Kingdom \label{inst58}
                \and Department of Physics \& Astronomy, University of Lethbridge, 4401 University Drive, Lethbridge, AB, T1K 3M4, Canada \label{inst59}
                \and Nicolaus Copernicus Astronomical Center, Polish Academy of Sciences, ul. Bartycka 18, 00-716 Warsaw, Poland \label{inst52}
                \and Steward Observatory and Department of Astronomy, The University of Arizona, Tucson, AZ 85721, USA \label{inst53}
                \and INAF - Osservatorio Astronomico di Padova, Vicolo dell'Osservatorio 5, 35122, Padova, Italy \label{inst55}
                \and Department of Physics and Astronomy, Vanderbilt University, Nashville, TN 37235, USA \label{inst56}
                \and School of Engineering and Applied Sciences, Department of Earth and Planetary Sciences, Harvard University \label{inst57}
                \and \href{https://www.LIFE-space-mission.com}{LIFE-space-mission.com}\label{instfinal}
              }

   \date{\today}

  \abstract
   {The increased brightness temperature of young rocky protoplanets during their magma ocean epoch makes them potentially amenable to atmospheric characterization to distances from the solar system far greater than thermally equilibrated terrestrial exoplanets, offering observational opportunities for unique insights into the origin of secondary atmospheres and the near surface conditions of prebiotic environments.}
    {The Large Interferometer For Exoplanets (LIFE) mission will employ a space-based mid-infrared nulling interferometer to directly measure the thermal emission of terrestrial exoplanets. Here, we seek to assess the capabilities of various instrumental design choices of the LIFE mission concept for the detection of cooling protoplanets with transient high-temperature magma ocean atmospheres at the tailend of planetary accretion. In particular, we investigate the minimum integration times necessary to detect transient magma ocean exoplanets in young stellar associations in the solar neighborhood.}
   {Using the LIFE mission instrument simulator (LIFEsim) we assess how specific instrumental parameters and design choices, such as wavelength coverage, aperture diameter, and photon throughput, facilitate or disadvantage the detection of protoplanets. We focus on the observational sensitivities of distance to the observed planetary system, protoplanet brightness temperature {using a blackbody assumption}, and orbital distance of the potential protoplanets around both G- and M-dwarf stars.}
   {Our simulations suggest that LIFE will be able to detect (S/N $\geq$ 7) hot protoplanets in young stellar associations up to distances of $\sim$100 pc from the solar system for reasonable integration times (up to $\sim$hours). Detection of an Earth-sized protoplanet orbiting a solar-sized host star at 1 AU requires less than 30 minutes of integration time. M-dwarfs generally need shorter integration times. The contribution from wavelength regions $<6$ \text{\textmu}m is important for decreasing the detection threshold and discriminating emission temperatures.}
   {The Large Interferometer for Exoplanets is capable of detecting cooling terrestrial protoplanets within minutes to hours in several local young stellar associations hosting potential targets. The anticipated compositional range of magma ocean atmospheres motivates further architectural design studies to characterize the crucial transition from primary to secondary atmospheres.}

   \maketitle
%

\section{Introduction}

The Large Interferometer For Exoplanets (LIFE) Collaboration (\href{https://www.LIFE-space-mission.com}{LIFE-space-mission.com}) develops the roadmap for the technical implementation and scientific exploitation of a spacebased midinfrared nulling interferometer to detect and characterize terrestrial exoplanets \citep{2022ExA....54.1197Q,quanz_et_al._2022}. Previous studies show promising simulation results for exoplanet detection and characterization focused on cool, potentially habitable exoplanets \citep{dannert_et_al._2022,2022A&A...664A..23K,2023A&A...673A..94K,2022A&A...665A.106A,2022A&A...668A..52K,2022A&A...664A..52H,2023A&A...670A..57H,2023AsBio..23..183A,2024AJ....167..128A,2023A&A...678A..96C,Kammerer_2018}, with possible variations in architectural and instrumentation design influencing the accessible parameter space \citep{2022A&A...664A..52H,2023A&A...670A..57H,2023A&A...678A..97M}. So far only a few studies have considered the extensive potential of the LIFE concept on questions surrounding the formation and evolution of planetary systems \citep{bonati_lichtenberg_bower_timpe_quanz_2019,2023A&A...671A.114J}.

In this manuscript, we focus on the potential of finding terrestrial protoplanets in their early, transient magma ocean phase following planetary accretion, when rocky planets are globally molten rather than solid. Highly molten magma ocean phases are a standard outcome of planetary formation \citep{2021ChEG...81l5735C,2023ASPC..534..907L}, since both planetesimal-based models \citep[e.g.,][]{Alibert2013,2016ApJ...821..126Q, Wyatt2019,Emsenhuber2021,Schlecker2021,Schlecker2021b,2020PSJ.....1...18C,2022ApJ...928...91C,2022ApJ...938L...3L} and pebble accretion models \citep[e.g.,][]{Morbidelli2012a,Lambrechts2019b,2021SciA....7..444J,2023A&A...671A..75J,2022E&PSL.58717537O,2023E&PSL.62218418O} predict a phase of high-energy impacts. This phase of planetary evolution is of crucial importance for the transition between primary and secondary atmosphere, as during highly molten phases the distribution of atmospheric volatiles between planetary metal core, silicate mantle, and volatile envelope is established \citep{2023ASPC..534..907L,2023ASPC..534.1031K,2023FrEaS..1159412S}.

Why do we need to observationally characterize magma ocean epochs for a better understanding of planetary evolution, rather than solely mature secondary atmospheres of terrestrial exoplanets? Earth-sized planets with secondary atmospheres are typically thought to host oxidizing planetary mantles due to the pressure sensitivity of internal geochemical reactions \citep{2019Sci...365..903A,2021SSRv..217...22G,2022GeCoA.328..221H,2023E&PSL.61918311H}, which means their volcanic emissions would be rich in gasses such as \ce{CO2}, \ce{H2O}, and \ce{SO2}, influencing longterm climatic and surface conditions.  However, whether the prebiotic climate on Earth at the transition between primary and secondary atmosphere was equally oxidizing is an ongoing enigma \citep{2020plas.book....3Z,2020SciA....6.1420C}, but crucially important for the chemical synthesis of surficial life \citep{2020SciA....6.3419S,benner2020did}. Placing temporal constraints on the transition of a planet from a primary atmosphere to a secondary atmosphere, and potentially its surface temperature regime, will allow us to better understand when the surface of early Earth (and other terrestrial rocky planets) becomes suitable for life as we know it. If one was to take the upper temperature limit of life to be 122$^\circ$C \citep{2008PNAS..10510949T}, observations from LIFE will allow us to better estimate the time between accretion and surface conditions suitable for life. Furthermore, the study of early atmospheric transitions will allow us to understand the early surface chemical conditions for life, including redox state and atmospheric gases available for metabolism, better allowing us to elucidate the physical and chemical conditions in which life emerges after accretion and the chemical inventory of molecules available to a potential biosphere in the earliest stages of its emergence \citep{2023NatAs.tmp..267C}. 

At least part of the answer to the chemical nature of the volatile envelope at this transitional stage lies in the chemical segregation that occurs during the magma ocean epoch \citep{2023FrEaS..1159412S,LichtenbergMiguel2024}. During these highly molten episodes, the chemical transfer of atmospheric volatiles between core, mantle, and atmosphere is rapid due to $\sim$hour to $\sim$day-long convection timescales in planetary magma oceans \citep{Solomatov_2015}. Vigorous convection in magma oceans diffuses the boundary between metallic core and silicate mantle \citep{2021ApJ...914L...4L}, which affects the composition of the outgassed atmosphere \citep{2022PSJ.....3..127S}. Dissolution of volatiles, such as water, into the magma ocean, and loss to space further alters the chemical composition of the magma ocean atmosphere \citep{2016ApJ...829...63S,2021JGRE..12606711L,2021ApJ...922L...4D,2022PSJ.....3...93B}. Therefore, being able to study the composition of magma atmospheres in extrasolar planetary systems would substantially aid in interpreting the scarce geologic record of the prebiotic Earth, and potentially improve our understanding of the conditions that enabled the origin of life as we know it \citep{2001PNAS...98.3666S,2020plas.book....3Z}. In addition, it would gain as valuable insights into the potential abundance of reduced exoplanet climates with variable greenhouse forcing and thus liquid water stability \citep{2011ApJ...734L..13P,2017ApJ...837L...4R}.

Furthermore, the study of terrestrial planets during their magma ocean phase will provide critical insight regarding the eventual pathway of volatile inventories, most particularly water content. \citet{2013Natur.497..607H} showed that the duration of the magma phase for terrestrial planets can determine whether water is able to condense on the surface and form oceans, or remain in the atmosphere as a greenhouse gas. Indeed, the climate simulations of \citet{2021Natur.598..276T} demonstrate the possibility that water remaining in the atmosphere can migrate to the night side of the planet, warming the surface further and preventing the ocean formation, possibly explaining the evolutionary pathway of Venus relative to Earth \citep{2019JGRE..124.2015K}. Alternatively, if water oceans were able to form on an early Venus, then cloud formation at the substellar point may have allowed temperate conditions to remain for several billion years \citep{2020JGRE..12506276W}. The resolution of these various scenarios will be critically important in a proper evaluation of terrestrial exoplanets found within the Venus Zone of their stars \citep{2014ApJ...794L...5K,2023AJ....166..199O} and understanding their evolutionary pathways.

The Large Interferometer For Exoplanets can also shed light on another important feature of planetary evolution, namely, the presence and influence of the primordial envelope. During the first few million years, the environment around the stars is dominated by a protoplanetary disk of gas and dust, often referred to as the solar nebula, in which the planets form. Disks likely survive 2–10 Myr \citep{2009AIPC.1158....3M}. Within the disk, a planet of sufficient mass can accrete a primordial envelope from the disk \citep{2015A&A...576A..87S,2016MNRAS.459.4088O,2020SSRv..216...74L}.  How rapidly this envelope is then lost depends on the evolution of the young star. Some of it might still be present in the modern Earth, having been dissolved in the molten mantle \citep{2019Natur.565...78W}. Contribution from this primordial envelope might influence the magma ocean composition and the observational fingerprints of the hot planets seen by LIFE.

Previous work on this stage of planetary evolution has focused on debris disks in exoplanetary systems and the detection of post-collision dust clouds as important circumstantial evidence of giant impacts at the tailend of planetary accretion \citep[e.g.,][]{2007ApJ...658..569W,2008ApJ...673.1123L,2012MNRAS.425..657J,2018MNRAS.479.2649K,2018AJ....155..194E,2019ApJ...875...45T,2019AJ....157..202S,2019MNRAS.488.4465G,2022MNRAS.514.1386G,2020MNRAS.497.2811K,2021AJ....161..186D,2023A&A...671A.114J,2023MNRAS.526.3115B,2023MNRAS.519.3257H,2023Natur.622..251K}. Here, in contrast, we focus on the detection of the magma ocean surface or magma ocean atmosphere itself \citep{1994AJ....108.2312S,2003ApJ...596L..95Z,2007ApJ...668L.175M,2009ApJ...704..770M,2014ApJ...784...27L,hamano_et_al._2015,bonati_lichtenberg_bower_timpe_quanz_2019}. In the literature this is sometimes termed {collision afterglow}, with the giant impact of terrestrial planetary accretion in mind. Here, however, we are agnostic of the physical nature of the heating event and solely focus on the potential for a significantly bright point source in exoplanetary systems in different observational circumstances. In general, magma ocean surfaces and atmospheres are assumed to be hot (1000-3000 K) and thus bright in the near and mid-infrared (IR) \citep{magma_oceans_2022}. However, with varying atmospheric composition, crystallization state over time \citep{2021JGRE..12606711L,2021PSJ.....2..207G} and steam dominated protoplanets typically settling quickly onto the Simpson-Nakajima limit \citep[the thermal emission threshold defining the inner edge of the classical habitable zone concept; e.g.,][]{2013NatGe...6..661G,2015AsBio..15..362G,2021ApJ...919..130B}, the emission temperature of protoplanets could be several hundreds of K lower. Thus we are considering protoplanets of a wide range of potential emission temperatures.

The work by \cite{bonati_lichtenberg_bower_timpe_quanz_2019} treated the observational prospects of LIFE's 5.6 \text{\textmu}m and 10 \text{\textmu}m filters to detect post-impact afterglows in nearby young stellar associations.  In this manuscript, we consider different instrument parameters of the LIFE concept comprising wavelength ranges that extend from a minimum of 3 \text{\textmu}m to 20 \text{\textmu}m. We evaluate the relative performances and analyze under which circumstances a cooling protoplanet can be detected by LIFE in a certain distance range and how much time these observations would require. Our goal is to study the instrument's capabilities and the possibility of detecting magma ocean protoplanets in young stellar associations.

\section{Methods}
\label{Methods}

\subsection{Instrument set-up and scenario}

For performance estimations of the LIFE instrument in detecting hot protoplanets we LIFE's mission simulator (LIFEsim) \citep{dannert_et_al._2022}, which simulates observations performed by the LIFE concept and allows parameter manipulation to study different instrumental setups. The LIFE simulator includes in its model the contribution of the major astrophysical noises, like photon noise and dust emission from the local and exozodiacal disks. {LIFEsim calculates each astrophysical noise term individually, these terms are then integrated over a two-dimensional artificial image. The distribution of exozodiacal dust is simulated according to the model by \citet{kennedy2015}.
The dust emission model in LIFEsim assumes that its distribution is similar to the solar system's. The emission is optically thin and LIFEsim approximates its (dimensionless) face-on surface density with a power law distribution \citep{dannert_et_al._2022}. The emission is then modeled as blackbody radiation from a face-on disk (temperature-radius profile). The disks are also assumed to always be face-on, symmetric and homogeneous. A study of the impact of angled disks on observations with LIFEsim can be found in \citet{quanz_et_al._2022}.}
LIFEsim does not yet include, however,  instrumental noise. We handle this uncertainty by adding an educated guess of the potential instrumental noise range to our calculations, further discussed below.

For this paper all but three instrument parameters were left to the default values from \citet{quanz_et_al._2022}, as summarized in Table \ref{tab:instrument_values}. The set of these three parameters constitutes the observing scenario for LIFE of which three major variations were considered, termed the \textsc{baseline}, \textsc{optimistic} and \textsc{pessimistic} scenarios. These scenarios were defined in \cite{quanz_et_al._2022} and can be found in Table \ref{scenarios}.

\begin{table}[h!]
\center

\caption{LIFEsim instrument setup and major parameters.}
\label{longlist}
\begin{tabular}{lll}
\hline
\textbf{PARAMETER} & \textbf{UNIT}         & \textbf{DEFAULT}     \\ \hline
\begin{tabular}[c]{@{}l@{}}Aperture diameter \end{tabular}                                     & {[}m{]}      & 2           \\
\begin{tabular}[c]{@{}l@{}}Minimum wavelength\end{tabular}                                    & {[}\text{\textmu}m{]} & 4           \\
\begin{tabular}[c]{@{}l@{}}Maximum wavelength\end{tabular}                                    & {[}\text{\textmu}m{]} & 18.5        \\
\begin{tabular}[c]{@{}l@{}}Quantum efficiency\end{tabular}                                    & {[}\%{]}     & 70          \\
\begin{tabular}[c]{@{}l@{}}Photon throughput \end{tabular}                                    & [\%]         & 5            \\
\begin{tabular}[c]{@{}l@{}}Spectral resolution\end{tabular}                                   &             & 20          \\
\begin{tabular}[c]{@{}l@{}}Minimum nulling length\end{tabular}                              & {[}m{]}      & 10          \\
\begin{tabular}[c]{@{}l@{}}Maximum nulling length\end{tabular}                              & {[}m{]}      & 100         \\
\begin{tabular}[c]{@{}l@{}}Imaging/Nulling length ratio\end{tabular}                        &       & 6           \\
\begin{tabular}[c]{@{}l@{}}Slewing time\end{tabular}                                          & {[}s{]}      & 36000         \\
\begin{tabular}[c]{@{}l@{}}Time efficiency\end{tabular}                                       & {[}\%{]}     & 80          \\ \hline
\begin{tabular}[c]{@{}l@{}}Image size\end{tabular} &                                      {[}pixel{]}           & 256         \\
\begin{tabular}[c]{@{}l@{}}Optimal wavelength \end{tabular}                                   & {[}\text{\textmu}m{]} & 15          \\ \hline
\begin{tabular}[c]{@{}l@{}}Local zodi model\end{tabular}                                       & -            & "darwinsim" \\
\begin{tabular}[c]{@{}l@{}}Habitable zone model\end{tabular}                                  & -            & "MS"        \\ \hline
\\
\end{tabular}
\label{tab:instrument_values}
\caption*{Notes: See \citet{dannert_et_al._2022} for the definition and detailed description of the parameters. The first 3 parameters change in the different observing scenarios described in Table \ref{scenarios}, the rest remain on \textsc{DEFAULT} value.}
\end{table}

\begin{table}[h!]
\center

\caption{The three observing scenarios for LIFEsim considered in this manuscript following \citet{quanz_et_al._2022}.}
\label{scenarios}
\begin{tabular}{llll}
\hline
Parameter               & \textsc{optimistic}    & \textsc{baseline}      & \textsc{pessimistic}   \\ \hline
Aperture (m)            & 3.5           & 2             & 1             \\
Min. $\lambda$ (\text{\textmu}m) & 3             & 4             & 6             \\
Max. $\lambda$ (\text{\textmu}m) & 20            & 18.5          & 17            \\ \hline
\\

\end{tabular}\\
\caption*{Notes: "Aperture" describes the aperture diameter of the collector spacecraft.}
\label{tab:scenarios}
\end{table}

The aperture diameter greatly affects the detection yield of the interferometer, especially when dealing with smaller, mature, planets. \citet{quanz_et_al._2022} found that the wavelength range increases the yield of only a few percent, while an aperture diameter of $D=3.5$ m resulted in an increase of around $100\%$ (averaged over different scenarios) with respect to an aperture of 2 meters. It is important to mention that throughout the simulations the photon throughput was kept fixed at $5\%$ \citep{quanz_et_al._2022}. This is a rather pessimistic prediction to ground the simulations in but positive results can build confidence for the instrument's capabilities, especially with a higher photon throughput (Appendix \ref{app:higherthroughput}).  We analyze the impact of these instrumental parameters differences when observing higher temperature protoplanets. Within this paper, as a simplifying assumption, the planets are treated as blackbodies of different emission temperatures. {Hot protoplanets will in fact show distinct emission features \citep{2009ApJ...704..770M,2014ApJ...784...27L}, however, the diversity expected for the variety of anticipated exoplanet compositions is large \citep{LichtenbergMiguel2024} and here we thus focus on scanning a larger parameter space with simplified assumptions. In Sect. \ref{subsec: limitations} we discuss this and future plans to build on this.}

When performing the simulations, we consider the different parameters stored in the data class of the simulator. Throughout the simulations we keep the planet's radius fixed at 1 Earth radius. For the planet's emission temperature we consider high temperature cases where the atmosphere is transparent to IR radiation, and moderate temperature cases where we assume the emission temperature is affected by the greenhouse effect of the magma ocean atmosphere. For the high temperature cases, we consider temperatures of $1500$ K, $2000$ K, $2300$ K, $3000$ K, $4000$ K, $7000$ K and $10000$ K, which are in the range of typically molten surfaces \citep{2023ASPC..534..907L}. The upper limit of $10^4$ K is an estimate for the emission temperature immediately after a giant impact event \citep{2018ApJ...861...52K,2020ApJ...901L..31K,2018JGRE..123..910L,2020E&PSL.53015885L}. For moderate emission temperatures, we use values inspired by radiating temperatures from \cite{2014ApJ...784...27L,hamano_et_al._2015,2021JGRE..12606711L,2021ApJ...919..130B}, with the effective temperatures of the planets ranging from $300$ K to $800$ K. {All planet spectra used in the simulations are blackbodies.}

\subsection{Signal-to-noise ratio and detection}
The output signal-to-noise ratio per wavelength bin, $(S/N)_{\lambda}$, is defined as the ratio between the photons detected from the exoplanet $S_{\lambda}$,  and the expected number of photons coming from the noise sources $N_{\lambda}$ \citep{dannert_et_al._2022}. We use Eq. (\ref{signall}) from \cite{dannert_et_al._2022} to obtain the total signal-to-noise ratio integrated over the entire considered wavelength range:

\begin{equation}
    (S/N)_{tot} = \sqrt{\sum_{\lambda}{\left(\frac{S}{N}\right)_{\lambda}}^2}.
\label{signall}
\end{equation}

\noindent Following \cite{dannert_et_al._2022}, we consider the detection threshold to be $(S/N)_{tot} \geq 7$, to compensate for the missing instrumental noise factor \citep{quanz_et_al._2022}.
 We assume that the instrumental noise contributes to the total noise less or equal to the astrophysical noise terms (see Eq. (\ref{eq: noise}) in Sect.\ref{subsec: limitations}). The threshold of $(S/N)_{tot} = 7$ corresponds to a signal-to-noise ratio of astrophysical sources (S/$N)_{astro} \geq 5$. {$2\sigma$ are added as an additional factor to account for the missing instrumental noise \citep[see the discussion in][]{dannert_et_al._2022}. In future, a different threshold shall be used once the impact of instrumental noise in LIFEsim is accurately quantified.} When the signal-to-noise ratio $(S/N)_{tot}$ of the simulation reaches 7 we can consider the instrument capable of detecting the planet's signal. This is what we consider as detection within the context of this manuscript. 

\subsection{Nearby young stellar associations}

Giant impacts that create magma ocean protoplanets typically take place in the early stage of a planetary system's lifetime. As a consequence of the decreasing number of planetesimals with time, the number of giant impacts around any star type decreases steeply with time, with most of them happening in the first 20 Myr after star formation \citep{2016ApJ...821..126Q,2021A&A...650A.152I,2022ApJ...928...91C,2022ApJ...938L...3L}. Because of the relation between stellar mass, number of giant impacts, and the size distribution of the resulting protoplanets \citep{bonati_lichtenberg_bower_timpe_quanz_2019}, the age and mass functions of the closest stellar associations are important factors when aiming to maximize the chances for detecting molten protoplanets \citep{2020ApJ...895..141B}. Therefore, while stellar associations containing a lot of M stars are more attractive when it comes to the likelihood of catching a planet in its magma ocean stage, these are typically smaller and hence less bright as their protoplanet counterparts around G and A stars, making them harder to detect  \citep{bonati_lichtenberg_bower_timpe_quanz_2019}. 

A list of known young stellar associations within 100 pc from the solar system, taken from \cite{bonati_lichtenberg_bower_timpe_quanz_2019}, can be found in Table \ref{associations}. We will use these as references when performing simulations and will examine a few of them further to evaluate the minimum integration times to detect cooling protoplanets around their stars. For the remainder of this manuscript, we will refer to the first 7 associations (in order of increasing distance) as the Near group, while the remaining three (32 Orionis, $\eta$ Chamaeleontis, and $\chi$ 1 For) as the Far group. This separation is due to a 30 pc gap between the two groups.

\begin{table}[h!]
\center

\caption{Young stellar associations with their respective distances to the geometric center of the group and age from \cite{Mamajek_2016}.}
\label{associations}
\begin{tabular}{lll}
Name                             & Distance [pc] & Age [Myr] \\ \hline
AB Doradus                       & 20            & 150       \\
\textbf{$\beta$ Pictoris}        & \textbf{37}   & \textbf{23}        \\
$\beta$ Tucanae                  & 43            & 45        \\
Tucana Horologium                & 48            & 45        \\
Columba                          & 50            & 42        \\
\textbf{TW Hydrae}               & \textbf{53}   & \textbf{10}        \\
Carina                           & 65            & 45        \\ \hline
32 Orionis                       & 92            & 22        \\
\textbf{$\eta$ Chamaeleontis}    & \textbf{94}   & \textbf{11}        \\
$\chi$ 1 For                     & 99            & 50   \\
\hline
\\

\end{tabular}\\
\caption*{Notes: For $\beta $ Pictoris' distance, the value found by \cite{bonati_lichtenberg_bower_timpe_quanz_2019} is used. Associations above the horizontal line are considered as {Near}, below the line as {Far}. The three associations studied in closer detail in this manuscript are highlighted in \textbf{bold}.}
\end{table}

\subsubsection*{M-type stars}
M-dwarf stars are of particular interest for detecting molten protoplanets because of their abundance relative to G-dwarf systems, among other considerations, such as their extended pre-main sequence phase. In \cite{quanz_et_al._2022}, simulations showed that the LIFE detection yield for planets around M-dwarfs was greater than for F, G and K stars due to their high occurrence rate in the solar neighborhood. Observing young stellar associations that contain M-dwarf stars also directly increases the likelihood of detecting cooling protoplanets with LIFE. It is important to note, though, that the previous survey simulations were performed for systems no further than 20 pc away, so we assess how the situation differs for systems at distances such as listed in Table \ref{associations}.

\noindent Given the wide range of masses and temperatures the M star spectral class covers, for the simulations we use the well-studied M3 star AD Leonis. It has an effective temperature of $3390$ K \citep{bárbara_rojas-ayala_covey_muirhead_lloyd_2012}, a radius of $0.4$ solar radii and a mass around $0.4$ solar masses \citep{reiners_gibor_basri_matthew_2009}. For achieving the same equilibrium temperature as Earth, a rapidly rotating planet with zero bond albedo around AD Leonis would orbit at about $0.134$ AU. We use this value to calculate orbits of interest for LIFE detection of terrestrial protoplanets. We remind the reader that we are interested in finding protoplanets with {transient} magma oceans, not atmosphere-stripped super-Earths or lava exoplanets, which are either permanently molten due to the intense irradiation \citep{2023ApJ...954...29P} or have lost both primary and secondary atmosphere, exposing the planetary interior dynamics to extreme stellar forcing \citep{2021ApJ...908L..48M,2023A&A...678A..29M}. Statistically, M-dwarf exoplanets are often found at these distances \citep{exostats}, hence no strong assumption is imposed.

Based on the above considerations, we pick three young stellar associations as main reference cases. These are: $\beta$ Pictoris, TW Hydrae and $\eta$ Chamaeleontis. $\beta$ Pictoris is considered a prime target for the detection of magma ocean exoplanets by \citet{bonati_lichtenberg_bower_timpe_quanz_2019}. Although AB Doradus is the closest association it may be too old to still contain magma ocean planets, thus $\beta$ Pictoris is the closest association that will be considered in the simulations. It is also considered as an observational target in \citet{2023A&A...671A.114J} because of its proximity and young age. The stellar association TW Hydrae is also of particular interest as it contains a lot of M stars among its population and is also particularly young, 10 Myr old. This, coupled with the fact that it is located at 53 pc away from Earth, makes TW Hydrae a very attractive candidate for detecting a cooling protoplanet in its magma ocean stage. Finally, $\eta$ Chamaeleontis is the only stellar association from the {Far} group that we study in greater detail. This again is due to its young age, far younger than the other two associations in the {Far} group (11 Myr as opposed to 22 and 50 Myr) and due to its high population of M stars. {M-dwarf stars in far-away systems (e.g. $\eta$ Chamaeleontis) can come close to the Inner Working Angle (IWA) of the instrument. See Sect. \ref{subsec: limitations} for a discussion of this.}

\section{Results}

\begin{figure*}[bth] 
\centering
\begin{center}
\includegraphics[width=0.85\textwidth]{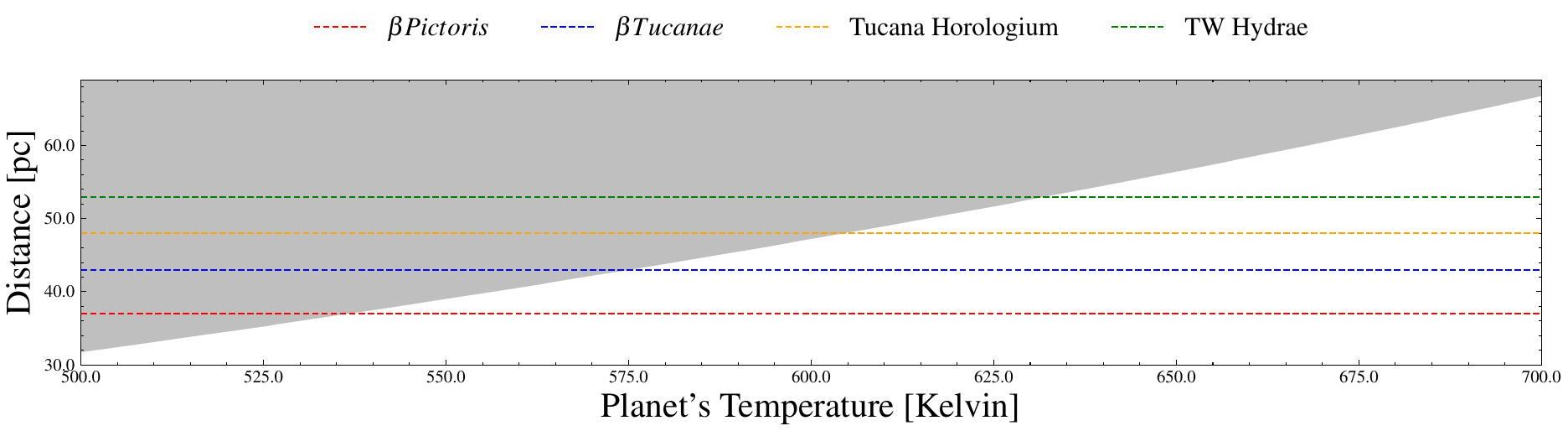}
\includegraphics[width=0.85\textwidth]{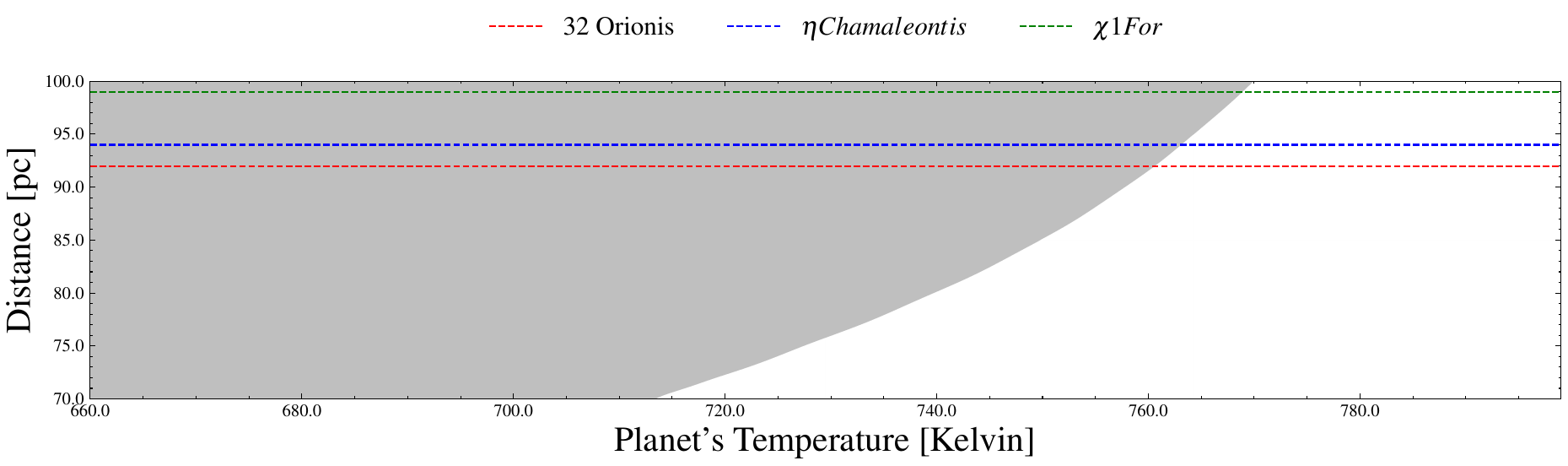}
\caption{Increase of maximum detectable distance with planetary emission temperature for an Earth-sized planet orbiting at 1 AU from a sun-like star, after 10 hours of integration time. The gray area corresponds to a S/N value below 7 (not detected) while the white is above 7 (detected). Several prominent young stellar associations are plotted as horizontal lines at their respective distances. {Top:} {Near} groups from Table \ref{associations}. {Bottom:} {Far} groups from Table \ref{associations}.}
\label{fig: 1}
\end{center}
\end{figure*}

\subsection{Moderate temperature cases}

Figure \ref{fig: 1} shows how the detection capabilities changes (white: detectable to above 7$\sigma$) as a function of emission temperature and distance, with several discussed young stellar associations indicated as dashed horizontal lines. The simulations are performed on an Earth-like planet orbiting a sun-like star at 1AU, with an integration time of 10 hours. Planets of temperature around $540$ K are already detectable in the closest considered stellar association $\beta$ Pictoris, while TW Hydrae's planets become detectable at around $635$ K. As the temperatures reach a threshold around $750$ K the increase becomes steeper as detection becomes relatively easier, {due to the planet's thermal emission becoming more significant}. The farthest considered stellar association ($\chi$ 1 For) reaches detection at around $770$ K. All considered associations within 100 pc become detectable below $800$ K.

\begin{figure}[h!] 
\includegraphics[width=0.45\textwidth, height=6.5cm]{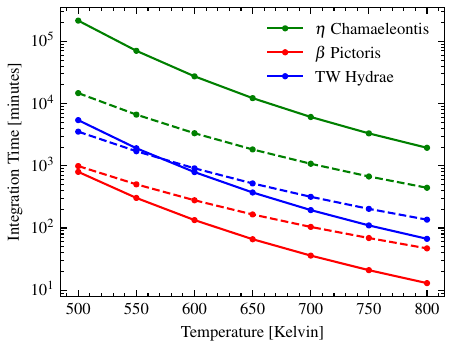}

\caption{{Moderate temperature cases}: Minimum integration times necessary for detection against the planet's emission temperature. The dashed lines represent the results for a solar-sized star host at 1 AU, while the solid lines are for M-dwarfs at 0.13 AU. The associations are at distances of 37 pc ($\beta$ Pictoris, red), 53 pc (TW Hydrae, blue) and 94 pc ($\eta$ Chamaeleontis, green).}
\label{fig: 2}
\end{figure}

The results were reached with a fixed integration time of 10 hours. It is clear that integration time is a very important factor which can alter the results considerably, so it is important to perform simulations to study its effect at these temperatures for the relevant stellar associations, which we consider in Fig. \ref{fig: 2}. As the distance to the sun of the star-planet system increases in Fig. \ref{fig: 2}, the integration time required for detection also increases. The distances are $37$ pc (red), $53$ pc (blue) and $94$ pc (green) and in all three cases the trends of the curve are almost identical. The integration time decreases exponentially as the temperature of the observed planet increases. 

For M-dwarf-orbiting planets the necessary integration times are consistently shorter for $\beta$ Pictoris. For TW Hydrae they are shorter for temperatures above 600 K, while for Chamaeleontis they are consistently much higher than their G star counterparts. {Targets are easier to detect around M stars at shorter distances but harder to detect at larger distances (with respect to targets orbiting G stars) because the targets orbiting M-dwarf stars at 0.13 AU fall within the IWA of LIFE in the farther association $\eta$ Chamaeleontis.} At shorter distances, for $\beta$ Pictoris, the times at 500 K are respectively almost 1000 minutes (16.6 h) for the sun-like star and 800 minutes (13.3 h) for the M-dwarf. Yet at $800$ K the times are much closer to each other, only 40 minutes apart.

\subsection{High temperature cases}

\begin{figure}[h!] 
\includegraphics[width=0.45\textwidth, height=6.5cm]{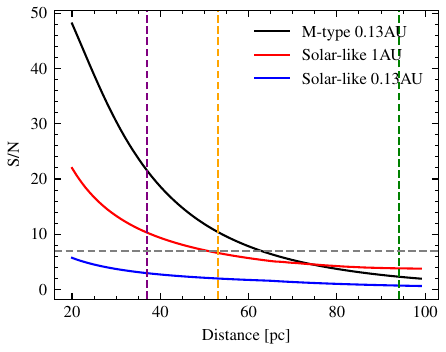}

\caption{S/N against distance for an Earth-sized 1500 K planet orbiting a solar-sized and an M-dwarf star. All results are after 5 minutes of integration time and the vertical dashed lines indicate the distance to the associations $\beta$ Pictoris (purple), TW Hydrae (orange), and $\eta$ Chamaeleontis (green), respectively. The horizontal gray dashed line represents the considered detection threshold (S/N = 7).}
\label{fig: 3}
\end{figure}

For protoplanet cases where the emission temperature is 1500 K or higher the situation is quite different (Fig. \ref{fig: 3}). The maximum detectable range in this case extends much farther. While the plot only shows results for distances up to 100 pc, these were obtained with a very short integration time of 5 minutes. The instrument can therefore reach detection at greater distances if enough time is employed.

At integration times of 10 hours, the results of Fig. \ref{fig: 1} indicate that the relation between distance and the planet's temperature is not linear. The distance reached after 10 hours at a temperature of $1500$ K is more than double compared to the distance at $750$ K. Furthermore,  Fig. \ref{fig: 1} shows that for reasonable integration times LIFEsim is able to detect a protoplanet in its magma ocean stage in the galactic neighborhood (<100 pc), where the considered stellar associations reside.

The S/N values from Fig. \ref{fig: 3} decrease with distance as farther objects are more difficult to observe. It is important to note that for LIFEsim it is easier to observe planets orbiting M-dwarfs at 0.13 AU than the same planet orbiting a solar-sized star at 1 AU, albeit for shorter distances, roughly up to 70 pc. Although the S/N values do converge at greater distances. With 5 minutes of integration time if the planet orbits a solar-sized star at 0.13 AU, LIFE would not be able to detect it even in $\beta$ Pictoris as the stellar leakage would make it impossible to observe (Fig. \ref{fig: 3}).

\subsubsection{G and M-dwarf star comparison}

\begin{figure}[h!] 
\includegraphics[width=0.45\textwidth, height=6.5cm]{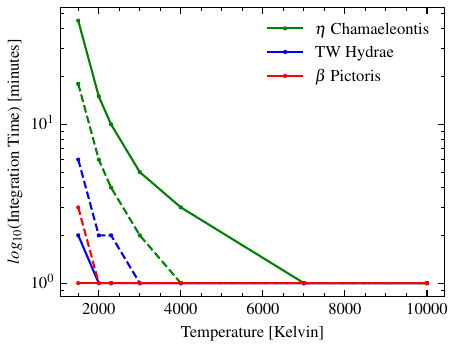}

\caption{{High temperature cases}: Minimum integration times necessary for detection against the planet's emission temperature. The dashed lines represent the results for a solar-sized star at 1 AU, while the solid lines are for M-dwarfs at 0.13AU. The associations are respectively at distances of 37 pc ($\beta$ Pictoris, red), 53 pc (TW Hydrae, blue) and 94 pc ($\eta$ Chamaeleontis, green).}
\label{fig: 4}
\end{figure}

Figure \ref{fig: 4} represents the minimum integration times necessary for detection of an Earth-sized planet for the {high temperature cases}. For the two closer associations, protoplanet detection for the M-dwarf cases at the respective orbits would theoretically only require one or two minutes of integration time. For $\eta$ Chamaeleontis the situation is very different with 45 minutes of integration time needed for the fainter planet. Results reach the 1 minute limit at around 7000 K. For the orbit around a solar-sized star the situation is different, with slightly higher times necessary (for $\beta$ Pictoris and TW Hydrae) at lower temperature but still only a few minutes are required. The lowest temperature planet reaches a maximum of 18 minutes in the farthest association $\eta$ Chamaeleontis.

\subsubsection{Varying LIFE parameters: solar-sized stars}

\begin{figure}[h!] 
\includegraphics[width=0.45\textwidth, height=6.5cm]{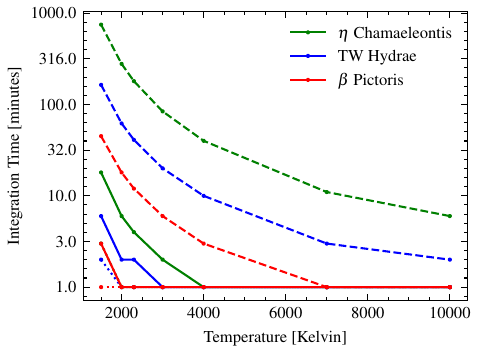}

\caption{{Varying LIFE parameters -- G star}:  Minimum integration times necessary for detection against the planet's emission temperature for an Earth-sized planet orbiting a solar-sized star at 1 AU. The simulations were performed in the \textsc{pessimistic} (dashed), \textsc{baseline} (solid), and \textsc{optimistic} (dotted) scenarios. The associations are respectively at distances of 37 pc ($\beta$ Pictoris, red), 53 pc (TW Hydrae, blue) and 94 pc ($\eta$ Chamaeleontis, green).}
\label{fig: 5}
\end{figure}

So far, all the results presented were simulated in the \textsc{baseline} scenario of LIFEsim. Next, we study the other observing scenarios and evaluate their performance and limitations. Fig. \ref{fig: 5} demonstrates a substantial difference between the different observing scenarios, for simulations on sun-like stars. In the \textsc{optimistic} scenario planets with temperatures of 2000 K or more only need 1 minute of integration time to be detected in all three stellar associations. In the \textsc{baseline} scenario, the three associations "flatline" at 1 minute, each at different temperatures, respectively at 2000, 3000 and 4000 K. The \textsc{pessimistic} scenario, as expected, is the one that requires the most integration time, with the closest association $\beta$ Pictoris reaching 1 minute at 7000 K. {The better performance of the optimistic scenario is due to its wider aperture diameter and wavelength range which allows for more signal to be collected from the targets.} The reason 1 minute is the minimum integration time is that we manually set it to that value. While the simulator allows for simulations down to one second, in a real setting, the instrument will not perform observations of such short times. In Fig. \ref{fig: 5} and Fig. \ref{fig: 6} the lines that are not visible are overlapping at the 1 minute level as they are all flat.

\subsubsection{Varying LIFE parameters: M-dwarf stars}
\label{sec: M-dwarf architecture}

\begin{figure}[h!] 
\includegraphics[width=0.45\textwidth, height=6.5cm]{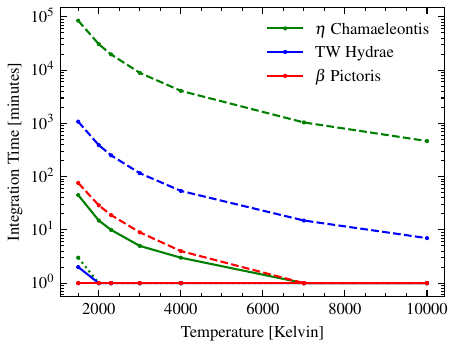}

\caption{{Varying LIFE parameters -- M-dwarf}:  Minimum integration times necessary for detection against the planet's emission temperature for an Earth-sized planet orbiting an M-dwarf at 0.13AU. The simulations were performed in the \textsc{pessimistic} (dashed), \textsc{baseline} (solid) and \textsc{optimistic} (dotted) scenarios. In the optimistic scenario all times are 1 minute. The associations are respectively at distances of 37 pc ($\beta$ Pictoris, red), 53 pc (TW Hydrae, blue) and 94 pc ($\eta$ Chamaeleontis, green).}
\label{fig: 6}
\end{figure}

 We perfom the same simulations, as in the previous section, for M-dwarfs. The results are displayed in Fig. \ref{fig: 6}. In the \textsc{pessimistic} scenario, $\beta$ Pictoris requires slightly more than an hour for the fainter planet (76 minutes), TW Hydrae requires almost 18 hours while Chamaeleontis requires 1416 hours. In the \textsc{baseline} scenario plot $\beta$ Pictoris is always 1 minute and TW Hydrae from 2000 K onward, while $\eta$ Chamaeleontis reaches it at 7000 K. In the \textsc{optimistic} scenario all three associations require 1 minute of integration time throughout the whole temperature range. Notably the integration time is below one hour in all cases.

\subsection{Influence of wavelength range on received total flux}

\begin{figure}[h] 
\includegraphics[width=0.45\textwidth, height=6.5cm]{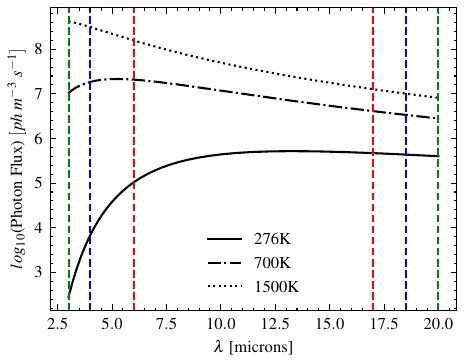}

\caption{Photon flux emitted by a blackbody against the wavelength (microns). Blackbody curves of 3 planets at $276$, $700$ and $1500$ K, respectively. The vertical lines represent the wavelength ranges considered for the \textsc{pessimistic} (red), \textsc{baseline} (blue) and \textsc{optimistic} (green) LIFE scenarios.}
\label{fig: 7}
\end{figure}

As one can see in Table \ref{scenarios}, the observing scenarios differ in two ways, the wavelength range considered and the aperture diameter of the collector spacecrafts. \citet{quanz_et_al._2022} found that the aperture diameter greatly affected the detection rate in their simulations, more significantly than the wavelength range. In the plots showcasing the integration times for the different scenarios (Fig. \ref{fig: 5} \& \ref{fig: 6}) it is clear that, while the \textsc{baseline} and \textsc{optimistic} scenarios are relatively similar to each other, the \textsc{pessimistic} scenario differs a lot from the other two. The \textsc{baseline} and \textsc{pessimistic} scenarios differ both in wavelength range and aperture diameter, with the aperture being $50\%$ of the \textsc{baseline} scenario. On top of this, the \textsc{baseline} and \textsc{optimistic} scenarios have different aperture diameters yet very similar results.

The reason for this disparity in the results also lies in the temperature of the observed planets. While \citet{quanz_et_al._2022} treated low temperature planets around the same effective temperature as the Earth, in this paper we deal with planets up to $10000 $ K. At higher temperatures, the emission starts to become increasingly dominated by shortwave radiation. This can be seen clearly in Fig. \ref{fig: 7}. For a 276 K planet, the flux ratio, integrated over all wavelengths, between the \textsc{pessimistic} and \textsc{baseline} scenario is of $88 \%$, while it is $82 \%$ for \textsc{pessimistic} over \textsc{optimistic}. On the other hand, for a 1500 K planet, the \textsc{pessimistic} over \textsc{baseline} ratio is of $40 \%$ and \textsc{pessimistic} over \textsc{optimistic} is at $24 \%$. This means that the \textsc{pessimistic} scenario receives only $40 \%$ of the photon flux from a 1500 K planet compared to the \textsc{baseline} scenario, and $24 \%$ of the \textsc{optimistic} one.

These results indicate that for the hotter planets we considered, the shorter wavelength regions become more important as they contain more photon flux from the observed planet. \citet{2021ApJ...919..130B} similarly found a shift in the spectrum toward shorter wavelengths (i.e. the visible regime) for correlated-$k$ radiation transport simulations of steam dominated atmospheres at runaway greenhouse conditions, which corroborates our findings despite the simplified blackbody assumptions.

\subsection{Constraining radiating temperature}
\label{subsec: constraining}

\begin{figure}[htb] 
\includegraphics[width=0.45\textwidth, height=6.5cm]{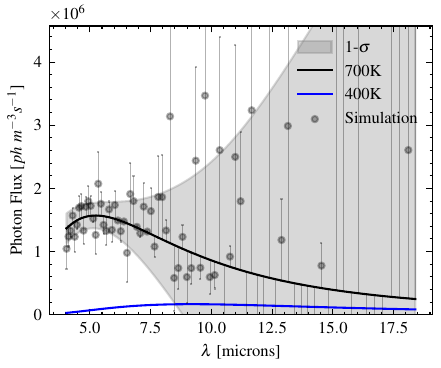}
\caption{Photon flux against wavelength comparing planets at 700 K (black) and 400 K (blue) emission temperature, respectively, with the \textsc{baseline} scenario parameters. The gray area represents the statistical difference of 1$\sigma$ from the 700 K planet. The scattered points are random noise to simulate the received flux from the warmer planet.}
\label{fig: 8}
\end{figure}

\begin{figure}[htb] 
\includegraphics[width=0.45\textwidth, height=6.5cm]{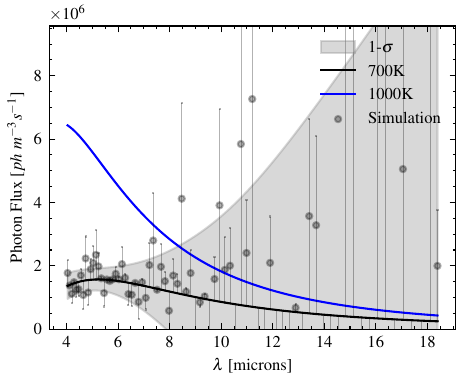}
\caption{Photon flux against wavelength comparing planets at 700 K (black) and 1000 K (blue) emission temperature, respectively, with the \textsc{baseline} scenario parameters. The gray area represents the statistical difference of 1$\sigma$ from the 700 K planet. The scattered points are random noise to simulate the received flux from the cooler planet.}
\label{fig: 9}
\end{figure}

In order to study how well LIFE could constrain a planet's temperature, we performed simulations on two planets with different temperatures and evaluated the resulting statistical difference. The simulations were performed for Earth-sized planets orbiting an M-dwarf star at 0.13AU. The distance to the system is 37 pc, as $\beta$ Pictoris. The cases studied in this paper are that of a 400 K and a 1000 K planet, each compared to a planet emitting at 700 K. The reason to simulate these cases is that the realistic emission temperature of a magma ocean planet would depend sensitively on its composition (redox state) and total volatile inventory, hence we test variations of several hundred K against each other. The uncertainty per bin was computed following

\begin{equation}
    \sigma_{bin} = \frac{F_{planet,bin}}{(S/N)_{bin}},
\end{equation}

where $F_{planet,bin}$ is the blackbody flux of the planet per wavelength bin and $(S/N)_{bin}$ is the signal-to-noise ratio per bin obtained by the instrument after the observation. The gray area in Fig. \ref{fig: 8} and Fig. \ref{fig: 9} represents the 1$\sigma$ statistical distance and the scattered points represent the blackbody flux with a random noise (between 0 and $\sigma_{bin}$) to simulate the received flux. The uncertainty is calculated for the standard 700 K planet, and the two plots represent the results when a +300 and -300 K planet's blackbody are plotted against this. For these simulations the spectral resolution of the instrument was set to 50.

\begin{figure}[htb] 
\includegraphics[width=0.45\textwidth, height=6.5cm]{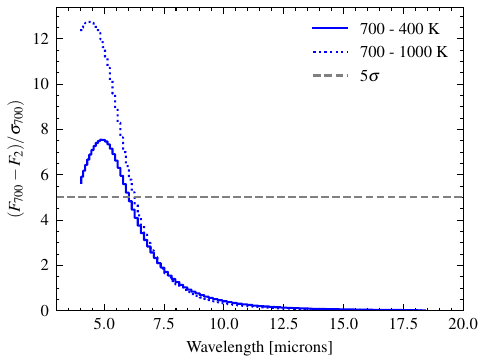}

\caption{Flux difference in units of $\sigma$ against wavelength for the two case studies of a 400 K or 1000 K compared against a 700 K protoplanet. F2 refers to the flux $F_{700}$ is compared against. The horizontal line marks the $5 \sigma$ threshold.}
\label{fig: 10}
\end{figure}

\begin{figure}[htb] 
\includegraphics[width=0.5\textwidth]{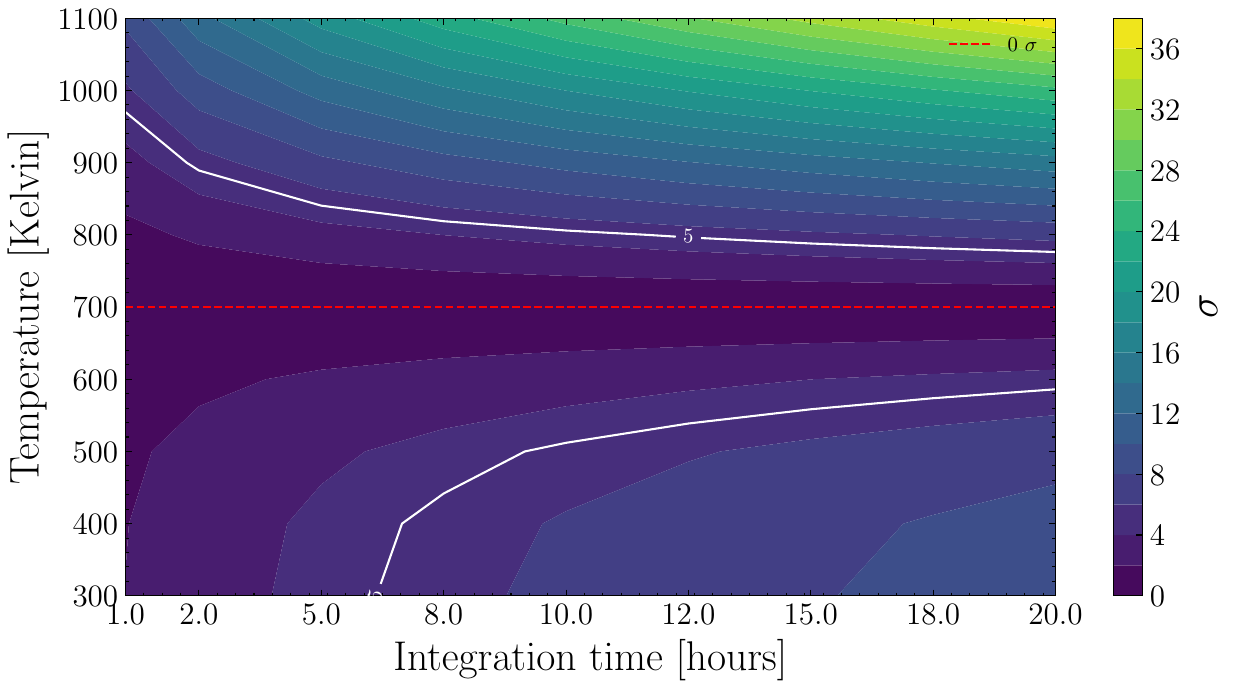}

\caption{Contour plot of the peak sigma value, from the statistical difference curve, for protoplanets of different emission temperatures against a 700 K planet.}
\label{fig: 11}
\end{figure}

Figures \ref{fig: 8} \& \ref{fig: 9} demonstrate further that the shorter wavelength regions become increasingly important for hotter emission temperatures. Below $\sim$10 \text{\textmu}m, the noise is low, but increases at longer wavelengths. Figure \ref{fig: 10} shows that the statistical difference between the planets in both cases are above the $5\sigma$ level only below $\lambda$ = 6 \text{\textmu}m. This means the LIFE mission would only be able to discriminate the emission temperatures in the significant climate regime for runaway greenhouse planets in this range for the \textsc{baseline} scenario (the \textsc{optimistic} would offer a wider range). {While the results of the analysis are shown per wavelength bin, the whole spectrum will be employed to determine the target planet's temperature. Therefore, regardless of the worse distinguishing power of longer wavelengths at these temperatures, an overall stronger distinction than $5\sigma$ between the two models will be reached.} In Fig. \ref{fig: 10} it is clear that while the two cases have the same temperature difference in magnitude ($300$ K), the result is more promising for the hotter planet case {as more signal is collected and the uncertainty per bin decreases}. The 700-1000 curve reaches a peak of almost 13 $\sigma$ while 700-400 reaches close to 8 $\sigma$. \\

Figure \ref{fig: 11} shows how the peak sigma value of the statistical difference between the planets varies with temperature difference and observation time. As expected, the greater the temperature difference and integration time, the higher the sigma value. The statistical difference increases faster for higher temperatures as the magnitude of the flux affects the observations, making it easier to distinguish fluxes for LIFEsim.

\subsection{Exozodi dust study}
\label{subsec: exozodi}

An additional factor that plays an important role in observations of young stellar systems is exozodiacal dust. It is expected that dust in young systems only tens of Myr old can reach very high levels compared to the solar system \citep{2010A&A...509A...9D,2014ApJ...780..154V,2017AstRv..13...69K,2021A&A...651A..45A}. For this reason, we performed simulations in the \textsc{baseline} scenario to study how exozodiacal dust affects LIFE's performance. Within these simulations we employ the exozodi dust prescription as introduced in \cite{dannert_et_al._2022}, hence we limit the exploration to dust belts similar to the terrestrial planet zone of the solar system. To keep the main body of the manuscript at a reasonable length we will discuss only 4 scenarios as per Fig. \ref{fig: 12} hereforth. A more extended parameter study can be can be found in \hyperlink{appendix1}{Appendix A}.\\

The contour plots in Fig. \ref{fig: 12} show that for the high temperature cases levels of 1000s of zodis do not pose a problem for detection. {The parameter "z" is a measure of the zodis of a system,  it is a unit that indicates the level of exozodiacal dust. One zodi (z=1) corresponds to the solar system level of zodiacal dust, z=2 is double and so on. It is therefore a global scaling factor.} The planet orbiting an M-dwarf obtains higher values for S/N consistently than its G-star counterpart, for example in the (d) plot the $S/N = 10$ line is outside the bounds of the plot. The results obtained for the 1500K planet indicate that even in the farthest association, with exozodiacal dust at levels thousands of zodis, LIFE is capable of detecting protoplanets. Realistically it will require times longer than the few minutes obtained in previous sections for z=3 but still reasonable times. For a more detailed study see \hyperlink{appendix1}{Appendix A}.\\

For the moderate temperature case of 700 K the situation is different. The detection limit for a sun-like orbiting planet at 1AU drops drastically, with detection becoming impossible for a thousand zodis even at 20 pc. See \hyperlink{appendix1}{Appendix A} for comparison with results after longer integration times. For M-dwarf-orbiting planets detection is still possible up to almost 70 pc for z=$10^3$, meaning that with a reasonably longer integration time, detection at the farthest association can be reached. \\

\begin{figure}[] 
\centering
\begin{center}
\subfloat[][]{\includegraphics[width=0.39\textwidth]{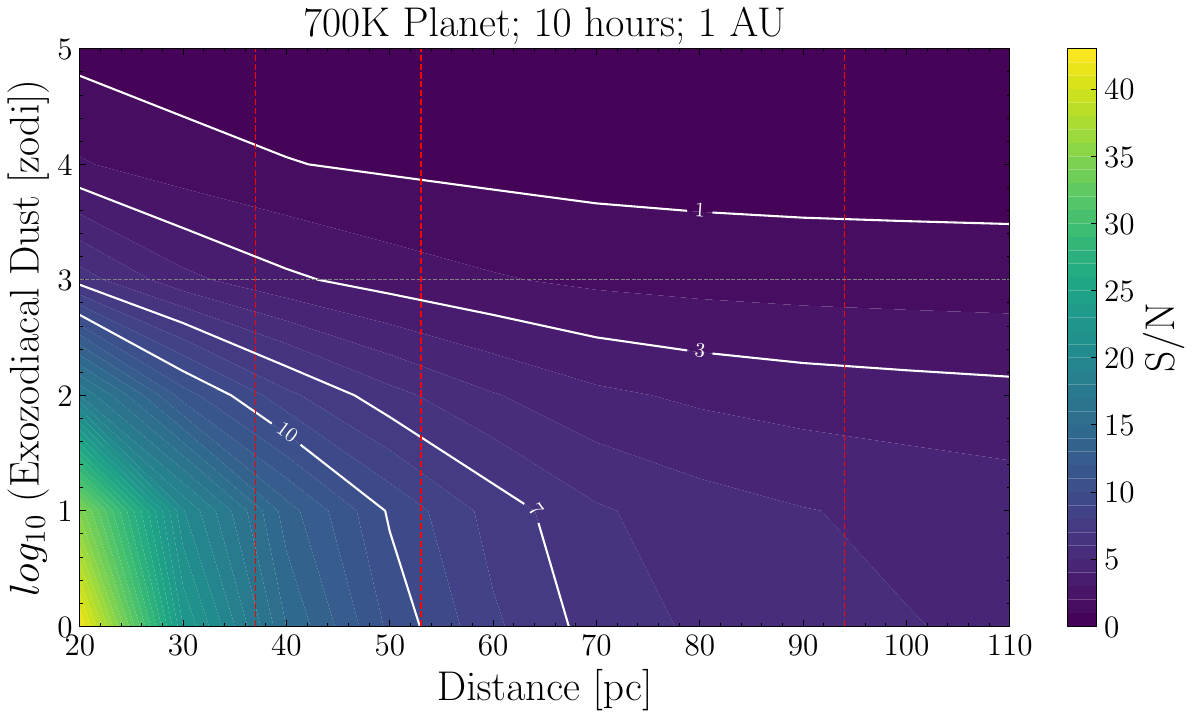}}\quad
\subfloat[][]{\includegraphics[width=0.39\textwidth]{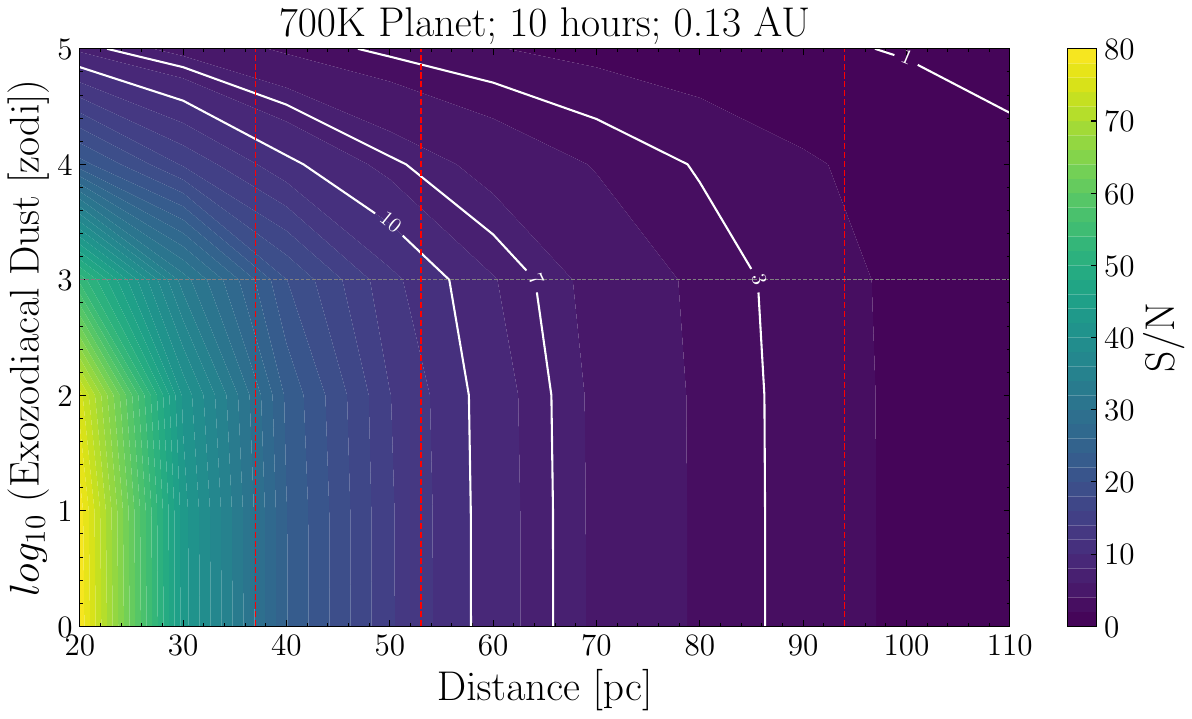}}\quad
\subfloat[][]{\includegraphics[width=0.39\textwidth]{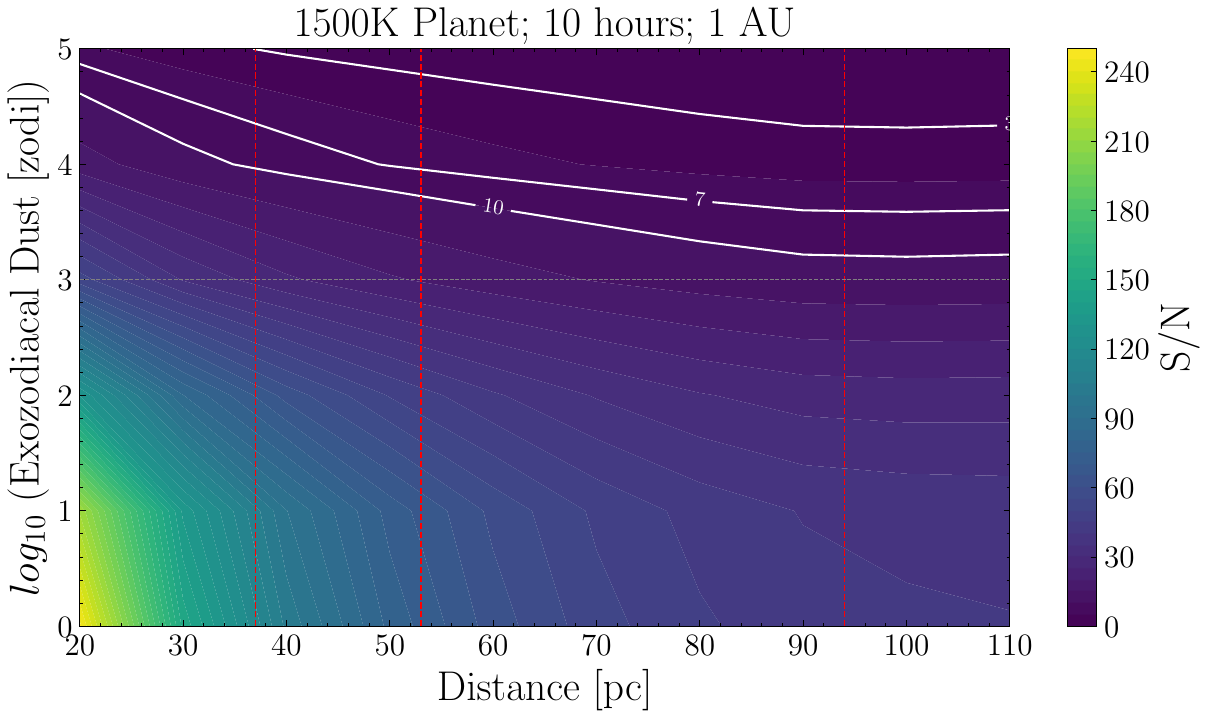}}\quad
\subfloat[][]{\includegraphics[width=0.39\textwidth]{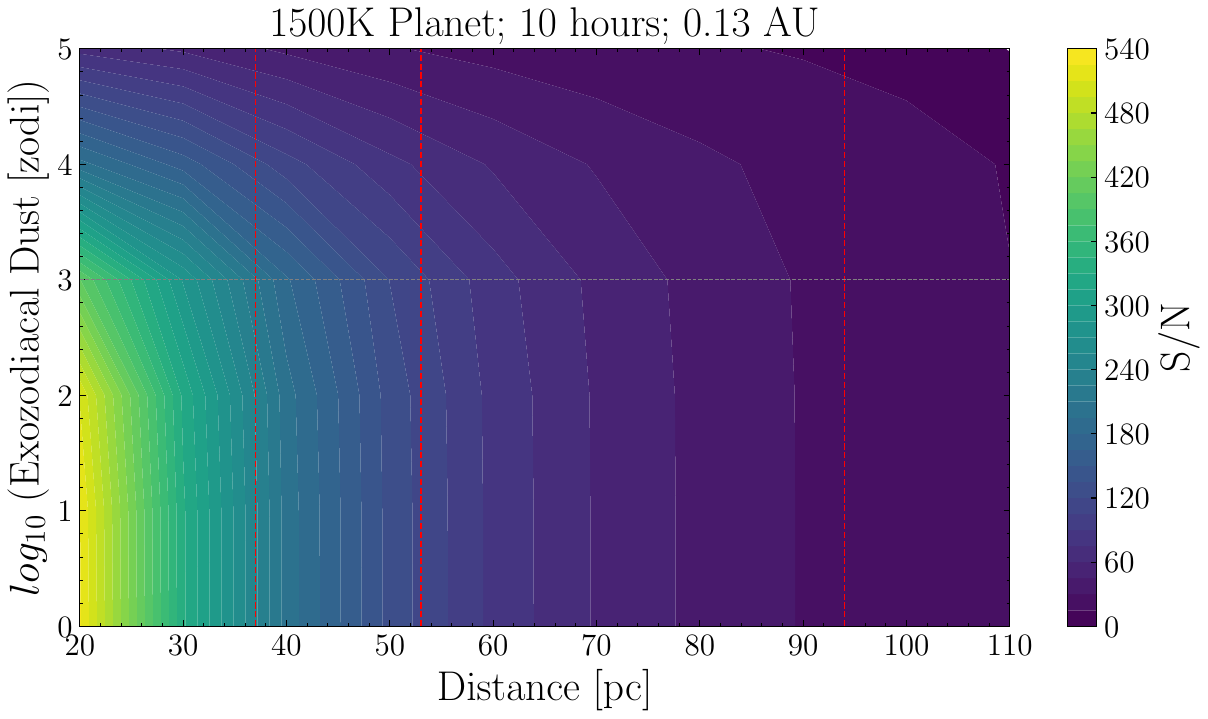}}\quad
\caption{Resulting S/N obtained after observation of a planet in a system with different exozodi dust levels. All four plots are obtained with 10 hours of integration time and for observation of a planet in the farthest stellar association $\eta$ Chamaeleontis. Plots (a) and (b) are observations for a moderate temperature case, 700 K, (c) and (d) are for a 1500 K scenario. For each temperature case the first plot is for a planet orbiting a sun-like star at 1 AU, the second is for an M-dwarf at 0.13 AU. The red vertical lines indicate the distances of $\beta$ Pictoris, TW Hydrae and $\eta$ Chamaeleontis, respectively.}
\label{fig: 12}
\end{center}
\end{figure}

\subsection{M-star varying luminosity}

\begin{figure}
    \centering
    \includegraphics[width=0.6\textwidth]{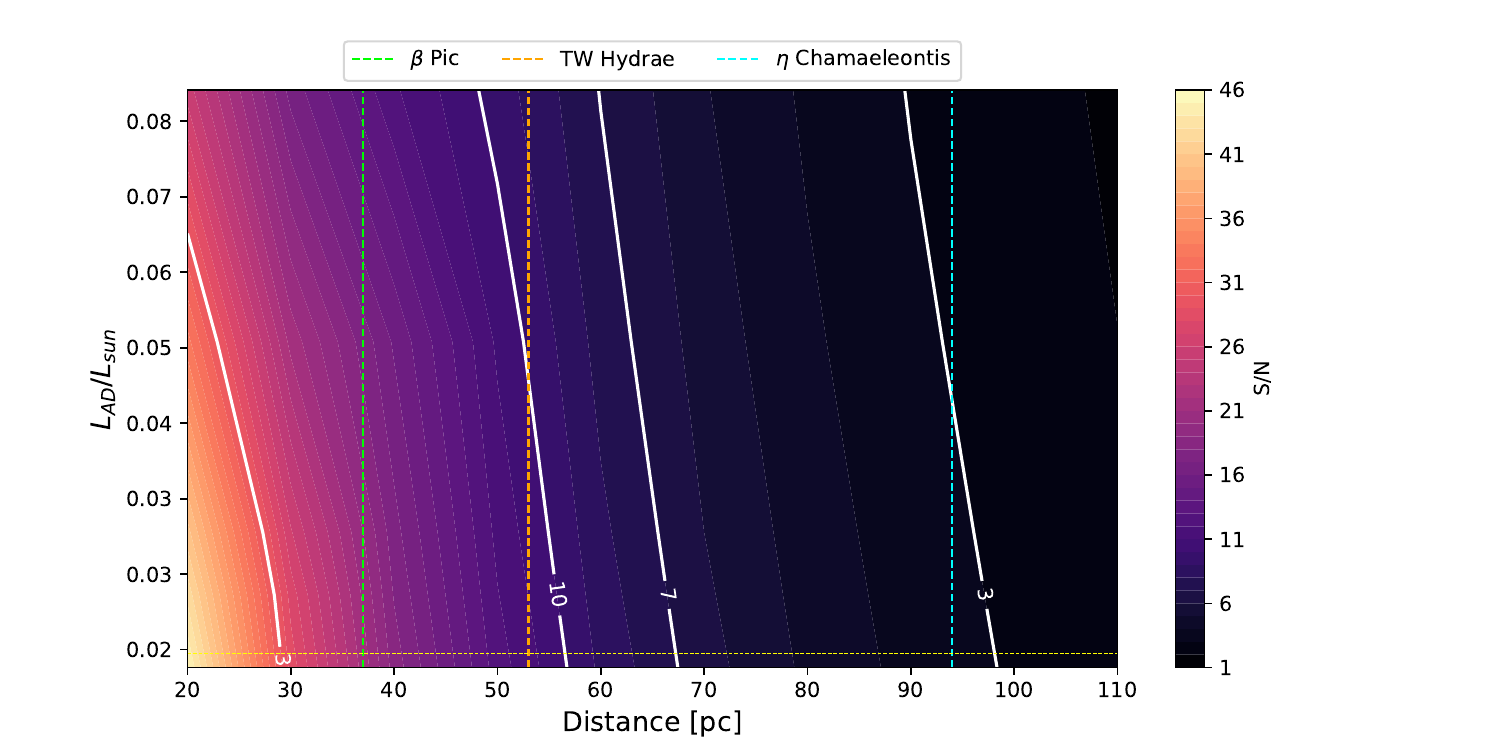}
    \caption{Countour plot of the S/N for a 1500 K planet orbiting AD Leonis M-dwarf star (orbit of 0.13 AU) at different luminosity (y-axis) and distance (x-axis), after one hour of integration time. The horizontal yellow line indicates the current AD Leonis to sun luminosity ratio.}
    \label{fig: 13}
\end{figure}

In the early stages of their lifetime M-dwarf stars are orders of magnitude brighter than their main sequence luminosity \citep{baraffe2015}. This paper focuses on M-dwarf stars in young stellar associations of the order of tens of Myr. Therefore it is important to explore how the luminosity of a star like AD Leonis affects simulations at different stages of its early life (see Fig. \ref{fig: 13}). The y-axis of Fig. \ref{fig: 13} represents the ratio of the luminosity of AD Leonis and the sun's luminosity, corresponding to different periods of the star's lifetime. The maximum of the luminosity corresponds to 10 Myr after star formation and the minimum at the current age luminosity \citep{baraffe2015}. The x-axis is distance to the system. In the earlier stages of its lifetime (high y-value) AD Leonis would be brighter than it is currently, therefore the ratio is higher. As time passes it decreases in luminosity until it reaches the horizontal yellow line which is the current AD Leonis luminosity to sun's luminosity ratio.

Figure \ref{fig: 13} shows that tracing back to the earliest stages of AD Leonis' lifetime the resulting detection distance varies by 5-6 pc at most. At its youngest stage, the simulation's detection capability reaches around 62-63 pc, while at the current stage it is at 67-68 pc at most. These results suggest that the fact that M-dwarf stars are brighter in the early stages of their lifetime will not significantly affect the observations of LIFE. Our results corroborate this for a 0.4 M$_\odot$ M-dwarf. For moderate temperature cases detection can still be reached at similar distances with reasonable integration times.

\section{Discussion}

\subsection{Distance range and integration times}

A major conclusion from our analysis is that the detection range for hot protoplanets of the LIFE concept extends in principle to a hundred pc, considerably beyond the local solar neighborhood. While the currently considered nominal ranges for potentially habitable planets for the LIFE survey are limited to nearby stars within 20 pc, the results obtained here provide strong support of considering observations further out in young stellar systems. Magma ocean atmospheres with emission temperatures in an intermediate range of a few hundred K are detectable after a reasonable integration time of 10 hours up to 100 pc and beyond. These results indicate that planets in the considered stellar associations can be detected with shorter integration times. In addition, a very high S/N for these planets can easily be obtained to characterize the magma ocean atmosphere in greater detail.

The main goals of this paper were to study LIFE's interferometer ranges and limits, considering different instrumental scenarios, when it comes to observing a magma ocean protoplanet. At present, the study of such hot atmospheres is focused on analyses of steam dominated runaway greenhouse climates \citep[e.g.,][]{2013NatGe...6..661G,2013Natur.504..268L,2015AsBio..15..362G,2015ExA....40..449L,2019Icar..317..583P,2021PSJ.....2..207G,2021ApJ...919..130B,2022A&A...658A..40C,2021Natur.598..276T}. During the early cooling phase after formation, these planets have brightness temperatures beyond $2000-3000$ K, with high temperature situations reaching up to $10000$ K immediately after collision events. Importantly, planets in steam dominated runaway greenhouse climate states will likely settle into an equilibrium point dictated by transmission through the two main water opacity windows \citep{2013NatGe...6..661G,2021ApJ...919..130B}. The equilibrium point in a runaway greenhouse climate is highly sensitive to the total water column and cloud distribution, and hence total atmospheric pressure, which makes observations of different planets in the runaway regime highly desirable for studying the deviation of volatile contents across planetary systems and the early climatic evolution of terrestrial protoplanets \citep{2022ARA&A..60..159W,2024PSJ.....5....3S}. In this paper we used LIFEsim to study the minimum integration times necessary to detect such scenarios with LIFE, with a focus on the young stellar associations $\beta$ Pictoris, TW Hydrae and $\eta$ Chamaeleontis. These associations are young (in the range of few Myr) as it makes it more likely to find young stars with magma ocean planets in their stars' orbit \citep{bonati_lichtenberg_bower_timpe_quanz_2019}.

There is a clear difference between observing solar-sized and M-dwarf stars, but for both cases the \textsc{baseline} and \textsc{optimistic} scenario stay below one hour integration time for the faintest considered protoplanet, among the high temperature cases, in the farthest stellar association. These results suggest that with reasonable amounts of observing time spent on these stellar associations there is a good chance of detecting a magma ocean protoplanet. Collecting more signal from a single planet is then comparably cheap in follow up studies, as a higher S/N can give more information on the object. Furthermore, the instrument's performance in the different observing scenarios showed interesting results. While \citet{quanz_et_al._2022} found that the aperture diameter affected the detection yield considerably more than the wavelength range (as for more temperate planets), the results obtained in this manuscript partly differ from this. It is worth pointing out that they considered a much wider sample in terms of both mass and age, focusing on mature exoplanets. The \textsc{pessimistic} scenario, which differs from the \textsc{baseline} scenario in both aperture diameter and wavelength range, displays much higher integration times. The \textsc{optimistic} and \textsc{baseline} have a 1.5 meters difference in aperture yet very similar results. The reason why the \textsc{pessimistic} scenario with its 1-m aperture diameter is so different can also, at least partly, be attributed to diminished wavelength range. These results indicate that restricting the wavelength range in the shortwave greatly affects the integration time necessary for the detection of molten protoplanets. The reason for this lies in the shift of the spectrum of hotter planets toward shorter wavelength ranges, compared to more temperate climate regimes.

Finally it is important to mention that, with regard to integration time, observational planning will need to consider the slewing time. In the LIFEsim environment, this time is set at 10 hours by default and it indicates the amount of time required to point the telescope toward a new target. This value will become more accurate in the future and varies depending on which targets it is moving between. The extremely short integration times per target star for reaching the detection threshold opens up the possibility of scanning essentially all young stars within the most ideal associations for the presence of terrestrial protoplanets. Magma ocean lifetimes of oxidized planets beyond the runaway greenhouse threshold are thought to last $\sim$Myr \citep{2017JGRE..122.1458S,2023Icar..39015265S,hamano_et_al._2015,Luger2015}, with potentially prolonging effects of primordial H-rich envelopes \citep{2021JGRE..12606711L}. The probability distribution of giant impacts beyond $\sim$10 Myr after stellar formation is strongly sensitive to the number of remaining planetary embryos at this stage \citep{2020PSJ.....1...18C,2022ApJ...928...91C}.

\subsection{G vs. M-dwarfs}

As outlined in \citet{quanz_et_al._2022}, LIFE shows a considerable positive bias toward finding exoplanets around M-dwarf stars, at least up to around 70 pc. Here, we further explored this effect by directly comparing the performances for solar-sized against M-dwarf stars. In this manuscript we considered orbital distances of $0.13$ AU for M-dwarfs. As a consequence of this, the smaller angular separation negatively affects simulations. M-dwarfs consistently require longer integration times for $\eta$ Chamaeleontis (94 pc) and for 500 and 550 K planets (Fig. \ref{fig: 2}) at TW Hydrae (53 pc). For the rest of simulations they perform better (or the same when overlapping at 1 minute) than their G star counterpart.  Furthermore the S/N trends in Fig. \ref{fig: 3} suggest that while it is better to observe planets around M-dwarf stars at shorter distances, the situation gets inverted after a certain point. As one can see in Fig. \ref{fig: 3} the curves intersect at approximately 75 pc. Beyond that, the S/N obtained from the planet orbiting a solar-sized star becomes greater. The reason for this intersection is likely related to the angular separation of the planet at those distances being very small and therefore requiring greater integration times. However, observing around M-dwarf stars fundamentally carries angular resolution as instrumental limitation. As previously discussed, when observing around M-dwarfs one needs to consider smaller orbits as those are the distances at which planets tend to form and orbit around these stars \citep{exostats}. See the extended discussion on angular separation limitations for M-dwarf planetary systems in Sect. \ref{subsec: limitations}.

Planets around M-dwarfs are anticipated to be locked in continuous runaway greenhouse phases for the first few hundred Myr of their lives because of the superluminous pre-main sequence phase of small stars \citep{2014ApJ...797L..25R,2015AsBio..15..119L,2022ARA&A..60..159W,2022ApJ...938L...3L}. This would give ample opportunity to observe magma ocean atmospheres in these planetary systems. The demographic trend of this extended runaway greenhouse phase should be observable already by transit missions \citep{2019A&A...628A..12T}, especially PLATO \citep{2024PSJ.....5....3S}. If detected statistically, LIFE will have substantial opportunity to study magma ocean atmospheres and climatic diversity if pointed toward a young M star planetary system. These investigations will also allow us to better understand the ability of M-dwarf orbiting planets to retain their atmospheres.

{Our work in this manuscript purely focuses on the simulated performance of the LIFE instrument in the detection and retrieval of hot protoplanets. The likelihood of detecting such objects in stellar systems is an important factor to take into account when considering a viable survey strategy. An initial study of the occurrence rates and timescales of hot protoplanets in young stellar systems has been performed in \cite{bonati_lichtenberg_bower_timpe_quanz_2019}, who they modeled the frequency of giant impact collisions using classical $N$-body planet formation models and the cooling of the resulting magma oceans, to calculate the likelihood of detecting "at least one magma ocean planet" in different stellar associations. They found that young stellar systems such as $\beta$ Pictoris and TW Hydrae are promising candidates for molten protoplanets detection with young stellar systems having higher occurrence rates of post-impact planetary objects. Our simulations here indicate shorter integration times necessary for detection, meaning that less observation time can be allocated to a single planet than was discussed in \cite{bonati_lichtenberg_bower_timpe_quanz_2019}. It is important to emphasize the actual number of giant impact events in young planetary systems across the stellar mass spectrum is subject to considerable uncertainty, in particular across the range of end-member planet formation scenarios from pebble- \citep[e.g.,][]{Morbidelli2012a,Lambrechts2019b,2021SciA....7..444J,2023A&A...671A..75J,2022E&PSL.58717537O,2023E&PSL.62218418O} to planetesimal-based \citep[e.g.,][]{Alibert2013,2016ApJ...821..126Q, Wyatt2019,Emsenhuber2021,Schlecker2021,Schlecker2021b,2020PSJ.....1...18C,2022ApJ...928...91C,2022ApJ...938L...3L} accretion frameworks. Therefore, the detection of molten protoplanets in such systems will serve to critically constrain the planetary accretion scenarios. A dedicated study to assess the survey trade-offs regarding slewing time between individual observations of young planetary systems and the predictions from end-member planet formation models should be conducted in future work. 
}
 
Throughout this paper we used a maximum nulling baseline of 100 m, similar to previous LIFE publications, which might be subject to change in the future. If one considers longer baselines, for instance, the maximum baseline of 168 meters assumed by \cite{bonati_lichtenberg_bower_timpe_quanz_2019}, observations become easier around M-dwarfs at 0.13 AU for $\beta$ Pictoris, TW Hydrae and $\eta$ Chamaeleontis. An increased nulling baseline would therefore increase the opportunity to detect and characterize protoplanets around M-dwarfs in these stellar associations. See Sect. \ref{subsec: limitations} for an extended discussion.

\subsection{Sensitivity to wavelength range}

We found that observing scenarios with a restricted wavelength range and aperture (e.g. the \textsc{pessimistic} scenario) perform worse than the other scenarios. Figure \ref{fig: 7} and the noise distribution in Figs. \ref{fig: 8}-\ref{fig: 9} suggest the shorter wavelengths regime rather than longer wavelengths as the reason for this discrepancy. By restricting the shorter wavelengths, the \textsc{pessimistic} scenario collects much less photon flux from the target planet. In the 1500 K case for example, we found that $24\%$ of the flux the \textsc{optimistic} scenario collects can also be collected by the \textsc{pessimistic} scenario, most of the missing flux lying in the short wavelengths (by several orders of magnitude). Furthermore, LIFE can distinguish between different emission temperatures better at lower wavelengths with $\sigma_{bin}$ becoming larger with increasing wavelength. For the purpose of constraining the effective temperature of a planet, wavelength regions $< 6$ \text{\textmu}m are more important than $> 17$ \text{\textmu}m.

\subsection{Limitations}
\label{subsec: limitations}

The results discussed in this manuscript are simulated, necessarily restricting the realism of the obtained noise estimations. The LIFE simulator includes all major astrophysical noise terms (e.g. zodiacal and exozodiacal dust, stellar leakage), which are always included in the simulations and accounted for in the final results. \cite{quanz_et_al._2022} and \cite{dannert_et_al._2022} studied the noise terms and their contributions in depth. They found that stellar leakage noise dominates in the short wavelength range, while at longer wavelengths the zodiacal dust becomes the main noise source. This is due to the fact that in the solar system the zodiacal dust is an important source of MIR emission, well in the wavelength range of the LIFE interferometer \citep{quanz_et_al._2022}. 

The LIFE simulator currently only includes the treatment of exozodiacal dust (see Appendix \ref{sec: appendixA}) for the photon noise from habitable zone dust.  It does not yet include a treatment of potential dust structures in the terrestial planet region that could cause confusion and hence mimic or hide the signal from a planet, neither does it include potential signals from hot dust closer to the star, which are observed around a fraction of stars \citep{2010A&A...509A...9D,2014ApJ...780..154V,2017AstRv..13...69K,2021A&A...651A..45A}. It will thus be important to incorporate a more elaborate treatment of exozodis to probe the potential for observing these systems with LIFE.

Furthermore, detector related noise and thermal background from the aperture mirrors and/or instrument optics are not included in the simulations yet. This is partly due to the fact that the instrument does not physically exist and the initiative is still considering different detectors and combiner architectures \citep{dannert_et_al._2022,Hansen_2022,2023A&A...670A..57H}. In this paper we dealt with this problem as in \cite{quanz_et_al._2022}, assuming that the instrumental noise contributes to the total noise less or equal to the astrophysical noise terms and that the total noise can be calculated using:
\begin{equation}
    \sigma_{tot} = \sqrt{(\sigma_{inst})^2 + (\sigma_{astro})^2}
    \label{eq: noise}
\end{equation}
\noindent where $\sigma_{inst}$ and $\sigma_{astro}$ are the instrumental and astrophysical noise terms respectively. By then choosing a S/N $\geq$ 7, corresponding to a S/$N_{astro} \geq 5$, as detection threshold, we hope to compensate for the lack of the instrumental noise term. A noise model is currently under implementation in LIFEsim. As of now, LIFEsim assumes that the shot noise (e.g. stellar leakage) will dominate over the instrumental noise \citep{dannert_et_al._2022}. Our results highlight the importance of shorter wavelength regimes for the LIFE mission, thus we need better modeling in these regimes where instrumental noise may dominate.

Especially for the more distant systems in our study, the inner region of the habitable zone tends to present at very small angular separations from the host star (e.g. the inner edge of the habitable zone around a $0.4 \, \mathrm{M_\odot}$ star in $\eta$ Chamaeleontis is at $1.2 \, \mathrm{mas}$). With the nulling baseline limited to 100 m this leads to these regions being well inside the typical Inner Working Angle (IWA), which is defined as the angular separation where the interferometric transmission drops to below 50\% ($\mathrm{IWA}_{10 \, \mathrm{\mu m}} = 5.2 \, \mathrm{mas}$). For these targets, LIFEsim correctly accounts for the low transmission. Close-in targets remain nonetheless detectable in photon count statistics because of their large midinfrared luminosity $L_\mathrm{bol} \sim T^4$ induced by their high effective temperatures. However, for distinguishing between the planetary and stellar signal, signal extraction in nulling interferometry relies on the modulation of the planetary signal by movement of the collector array \citep[see][]{dannert_et_al._2022}. In the assumed setup, the signals of planets well within the IWA receive a relatively low number of modulations per array rotation, which in turn can complicate signal extraction, especially in the presence of low frequency instability noise \citep{Lay04,Lay06}. This can be mitigated by allowing for longer nulling baselines. We plan to explicitly address this in future work by running detailed signal extraction on distant forming terrestrial exoplanets. 

From a target perspective, all our simulations use idealized blackbodies to simulate protoplanets of different emission temperatures. This is a substantial simplification, as atmospheres of varying compositions and emission temperatures differ greatly in their spectra. However, at high temperatures and pressures in the regime studied here, current climate simulations predict the emission to start behaving like a blackbody again \citep{2021ApJ...919..130B,2021JGRE..12606711L,2022A&A...658A..40C,2023Natur.620..287S}. This is, to some extent, rooted in the missing laboratory data on high temperature opacities \citep{2019astro2020T.146F}. Given the current uncertainties regarding high temperature and high pressure atmospheres over a wide composition array, our approach yields initial insights into the potential for observation. {This limitation also touches on the principal degeneracy in Sect. 3.4 between planetary emission temperature and surface. Typically both are degenerate with each other, in particular at short wavelengths. However, in the mid-IR this degeneracy can be broken because deeper regions of the atmosphere can be accessed \citep[e.g.][]{2022ARA&A..60..159W,2023A&A...673A..94K}. Therefore, we plan to significantly extend the current study, which focused on the detection of hot protoplanets, by simulating magma ocean atmospheres with more realistic climate models \citep[e.g.][]{2018AJ....155..195W,2023AN....34430075J}, to capture their anticipated diversity \citep{2023FrEaS..1159412S,LichtenbergMiguel2024,Kempton2023,Kempton2024} and assess the level of detail that can be retrieved.}

\section{Conclusions}

In this paper we investigated the range and required integration times for the detection of hot terrestrial protoplanets with the LIFE mission concept. We considered different parameters for the potential protoplanets and host stars focused on the planet's and star's temperature and radii. Additionally, we studied the detection capabilities of LIFE for protoplanets around M-dwarfs and evaluated the performance of different instrumental scenarios of LIFE, and their potential for discriminating the emission temperature of a given protoplanet. Our main findings can be summarized as follows:
\begin{itemize}
    \item The detection range for finding terrestrial protoplanets with LIFE extends far beyond the direct solar neighborhood, potentially reaching distances up to a hundred pc. A number of potentially interesting young stellar associations sit at distances ranging from 20 up to 95 pc. The Large Interferometer For Exoplanets can detect an Earth-sized magma ocean planet with integration times of minutes to hours at these distances.
    \item Detecting protoplanets with LIFE is generally easier for M-dwarf host stars. The simulations mostly showed better results (higher S/N and lower integration times) within this spectral type domain at distances below $\sim$70 pc, compared to G stars.
    \item The angular resolution of the instrument negatively affects the detectable orbital distance of terrestrial protoplanets past the direct solar neighborhood for M star targets. However, the anticipated performances of all considered LIFE scenarios still allow for detection within the terrestrial planet forming zone, in particular during the pre-luminous main sequence phase of M-dwarfs, which increases the likelihood of runaway greenhouse phases and thus magma ocean atmospheres.
    \item The performance of different instrumental design choices were evaluated and we found the \textsc{pessimistic} scenario to be more distinct from the other two analyzed scenarios. Our results indicate that the wavelength range being restricted in the shortwave affects the instrument's performance just as much, if not even more, than the aperture diameter. The shorter wavelength range is important when observing hotter planets because of the increase in shortwave emission at these temperatures.
    \item Because of this, discriminating the emission temperature of hot protoplanets with LIFE works better at shorter wavelengths. The resulting statistical difference between the observed planets passed the $5\sigma$ detection threshold only at $<$6 \text{\textmu}m. {Nonetheless when integrating over the whole spectrum, the resulting distinction will be greater than 5$\sigma$.}
    \item {The main limitations of our study are related to (i) the expected non-blackbody nature of protoplanet atmospheres, (ii) the limits to the inner working angle of LIFE for far-away systems, and (iii) the lack of an instrumental noise model. These limitations will be addressed in future work.}
\end{itemize}
  
In summary, the LIFE mission concept is capable of detecting and characterizing bright cooling protoplanets up to 100 pc. While these distances are not considered in the main target range for LIFE's core tasks, our results encourage observing targets in distinct young stellar associations at far greater distances from the solar system. Typically LIFE requires integration times of only minutes to a few hours for initial detection of protoplanet candidates. Characterizing terrestrial protoplanets would greatly contribute to our understanding of atmospheric formation on terrestrial planets, and provide detailed insight into the physical and chemical processes that occur during the transition from primary to secondary atmospheres, elucidating the conditions available for prebiotic chemistry on young worlds.

\section*{Data Availability}
\noindent 
The computer code to reproduce the results of this manuscript is openly available at \url{https://github.com/FormingWorlds/LIFE_MagmaOceanDetection}

\begin{acknowledgements}

The authors thank the anonymous reviewer and Daniel Angerhausen and Alycia Weinberger for comments and discussions that improved the paper.

T.L. acknowledges support from the Branco Weiss Foundation and the Alfred P. Sloan Foundation (AEThER project,  G202114194).
Parts of this work have been carried out within the framework of the National Centre of Competence in Research PlanetS supported by the Swiss National Science Foundation under grants 51NF40\_182901 and 51NF40\_205606.
The results reported herein benefited from collaborations and/or information exchange within NASA’s Nexus for Exoplanet System Science (NExSS) research coordination network sponsored by NASA’s Science Mission Directorate under Agreement No. 80NSSC21K0593 for the program ``Alien Earths".
S.P.Q. acknowledges the financial support of the SNSF. 
A.Fo. acknowledges support from the Swiss Space Office through the ESA PRODEX program. 
A.J. thanks the Danish National Research Foundation (DNRF Chair Grant DNRF159) and the Carlsberg Foundation (Semper Ardens: Advance grant FIRSTATMO).
D.D. acknowledges the support from the European Research Council (ERC) under the European Union's Horizon 2020 research and innovation program (grant agreement CoG - 866070).
V.S. acknowledges the support of the European Research Council (ERC) under the European Union’s Horizon 2020 research and innovation program (COBREX; grant agreement no. 885593).
C.C. acknowledges support from the Science and Technology Facilities Council (Grant  ST/Y001788/1). 
J.L.-B. is funded by the Spanish MICIU/AEI/10.13039/501100011033 and NextGenerationEU/PRTR grants PID2019-107061GB-C61 and CNS2023-144309.
A.C.-G. is funded by the Spanish Ministry of Science through MCIN/AEI/10.13039/501100011033 grant PID2019-107061GB-C61. 
J.K. gratefully acknowledges the support of the Swedish Research Council  (VR: Etableringsbidrag 2017-04945)
O.B.-R. is supported by INTA grant PRE-MDM-07, Spanish MCIN/AEI/10.13039/501100011033 grant PID2019-107061GB-C6, and CNS2023-144309. 
R.K. acknowledges financial support via the Heisenberg Research Grant funded by the Deutsche Forschungsgemeinschaft (DFG, German Research Foundation) under grant no.\~KU 2849/9, project no.~445783058.
This work has received funding from the Research Foundation - Flanders (FWO) under the grant number 1234224N.
G.C. thanks the Swiss National Science Foundation for financial support under grant number P500PT\_206785
RR edited the manuscript and provided feedback and analysis of results. 
M.T. is supported by JSPS KAKENHI grant No.18H05442.
M.B. acknowledges funding from the European Union H2020-MSCA-ITN-2019 under grant agreement no. 860470 (CHAMELEON).
S.K. acknowledges support by the European Research Council (ERC Consolidator grant, No. 101003096).
A.B.K. is supported by the Ministry of Science, Technological Development, and Innovation of R. Serbia through projects of  the University of Belgrade - Faculty of Mathematics (contract 451-03-68/2023-14/200104).

\end{acknowledgements}

\bibliographystyle{aa}
\bibliography{Ref}

\begin{thebibliography}{132}
\expandafter\ifx\csname natexlab\endcsname\relax\def\natexlab#1{#1}\fi

\bibitem[{{Absil} {et~al.}(2021){Absil}, {Marion}, {Ertel}, {Defr{\`e}re}, {Kennedy}, {Romagnolo}, {Le Bouquin}, {Christiaens}, {Milli}, {Bonsor}, {Olofsson}, {Su}, \& {Augereau}}]{2021A&A...651A..45A}
{Absil}, O., {Marion}, L., {Ertel}, S., {et~al.} 2021, \aap, 651, A45

\bibitem[{{Alei} {et~al.}(2022){Alei}, {Konrad}, {Angerhausen}, {Grenfell}, {Molli{\`e}re}, {Quanz}, {Rugheimer}, {Wunderlich}, \& {LIFE Collaboration}}]{2022A&A...665A.106A}
{Alei}, E., {Konrad}, B.~S., {Angerhausen}, D., {et~al.} 2022, \aap, 665, A106

\bibitem[{Alibert {et~al.}(2013)Alibert, Carron, Fortier, Pfyffer, Benz, Mordasini, \& Swoboda}]{Alibert2013}
Alibert, Y., Carron, F., Fortier, A., {et~al.} 2013, \aap, 558, 1

\bibitem[{{Angerhausen} {et~al.}(2023){Angerhausen}, {Ottiger}, {Dannert}, {Miguel}, {Sousa-Silva}, {Kammerer}, {Menti}, {Alei}, {Konrad}, {Wang}, {Quanz}, \& {LIFE Collaboration}}]{2023AsBio..23..183A}
{Angerhausen}, D., {Ottiger}, M., {Dannert}, F., {et~al.} 2023, Astrobiology, 23, 183

\bibitem[{{Angerhausen} {et~al.}(2024){Angerhausen}, {Pidhorodetska}, {Leung}, {Hansen}, {Alei}, {Dannert}, {Kammerer}, {Quanz}, {Schwieterman}, \& {The LIFE initiative}}]{2024AJ....167..128A}
{Angerhausen}, D., {Pidhorodetska}, D., {Leung}, M., {et~al.} 2024, \aj, 167, 128

\bibitem[{{Armstrong} {et~al.}(2019){Armstrong}, {Frost}, {McCammon}, {Rubie}, \& {Boffa Ballaran}}]{2019Sci...365..903A}
{Armstrong}, K., {Frost}, D.~J., {McCammon}, C.~A., {Rubie}, D.~C., \& {Boffa Ballaran}, T. 2019, Science, 365, 903

\bibitem[{Baraffe {et~al.}(2015)Baraffe, Homeier, Allard, \& Chabrier}]{baraffe2015}
Baraffe, I., Homeier, D., Allard, F., \& Chabrier, G. 2015, \aap, 577, A42

\bibitem[{Benner {et~al.}(2020)Benner, Bell, Biondi, Brasser, Carell, Kim, Mojzsis, Omran, Pasek, \& Trail}]{benner2020did}
Benner, S.~A., Bell, E.~A., Biondi, E., {et~al.} 2020, ChemSystemsChem, 2, e1900035

\bibitem[{{Bonati} {et~al.}(2019){Bonati}, {Lichtenberg}, {Bower}, {Timpe}, \& {Quanz}}]{bonati_lichtenberg_bower_timpe_quanz_2019}
{Bonati}, I., {Lichtenberg}, T., {Bower}, D.~J., {Timpe}, M.~L., \& {Quanz}, S.~P. 2019, \aap, 621, A125

\bibitem[{{Bonsor} {et~al.}(2023){Bonsor}, {Wyatt}, {Marino}, {Davidsson}, {Kral}, \& {Thebault}}]{2023MNRAS.526.3115B}
{Bonsor}, A., {Wyatt}, M.~C., {Marino}, S., {et~al.} 2023, \mnras, 526, 3115

\bibitem[{{Bottrill} {et~al.}(2020){Bottrill}, {Haigh}, {Hole}, {Theakston}, {Allen}, {Grimmett}, \& {Parker}}]{2020ApJ...895..141B}
{Bottrill}, A.~L., {Haigh}, M.~E., {Hole}, M. R.~A., {et~al.} 2020, \apj, 895, 141

\bibitem[{{Boukrouche} {et~al.}(2021){Boukrouche}, {Lichtenberg}, \& {Pierrehumbert}}]{2021ApJ...919..130B}
{Boukrouche}, R., {Lichtenberg}, T., \& {Pierrehumbert}, R.~T. 2021, \apj, 919, 130

\bibitem[{{Bower} {et~al.}(2022){Bower}, {Hakim}, {Sossi}, \& {Sanan}}]{2022PSJ.....3...93B}
{Bower}, D.~J., {Hakim}, K., {Sossi}, P.~A., \& {Sanan}, P. 2022, \psj, 3, 93

\bibitem[{{Carri{\'o}n-Gonz{\'a}lez} {et~al.}(2023){Carri{\'o}n-Gonz{\'a}lez}, {Kammerer}, {Angerhausen}, {Dannert}, {Garc{\'\i}a Mu{\~n}oz}, {Quanz}, {Absil}, {Beichman}, {Girard}, {Mennesson}, {Meyer}, {Stapelfeldt}, \& {LIFE Collaboration}}]{2023A&A...678A..96C}
{Carri{\'o}n-Gonz{\'a}lez}, {\'O}., {Kammerer}, J., {Angerhausen}, D., {et~al.} 2023, \aap, 678, A96

\bibitem[{{Catling} \& {Zahnle}(2020)}]{2020SciA....6.1420C}
{Catling}, D.~C. \& {Zahnle}, K.~J. 2020, Sci Adv, 6, eaax1420

\bibitem[{{Chao} {et~al.}(2021){Chao}, {deGraffenried}, {Lach}, {Nelson}, {Truax}, \& {Gaidos}}]{2021ChEG...81l5735C}
{Chao}, K.-H., {deGraffenried}, R., {Lach}, M., {et~al.} 2021, Chemie der Erde / Geochemistry, 81, 125735

\bibitem[{{Chaverot} {et~al.}(2022){Chaverot}, {Turbet}, {Bolmont}, \& {Leconte}}]{2022A&A...658A..40C}
{Chaverot}, G., {Turbet}, M., {Bolmont}, E., \& {Leconte}, J. 2022, \aap, 658, A40

\bibitem[{{Clement} {et~al.}(2020){Clement}, {Kaib}, \& {Chambers}}]{2020PSJ.....1...18C}
{Clement}, M.~S., {Kaib}, N.~A., \& {Chambers}, J.~E. 2020, \psj, 1, 18

\bibitem[{{Clement} {et~al.}(2022){Clement}, {Quintana}, \& {Quarles}}]{2022ApJ...928...91C}
{Clement}, M.~S., {Quintana}, E.~V., \& {Quarles}, B.~L. 2022, \apj, 928, 91

\bibitem[{{Cockell} {et~al.}(2023){Cockell}, {Simons}, {Castillo-Rogez}, {Higgins}, {Kaltenegger}, {Keane}, {Leonard}, {Mitchell}, {Park}, {Perl}, \& {Vance}}]{2023NatAs.tmp..267C}
{Cockell}, C.~S., {Simons}, M., {Castillo-Rogez}, J., {et~al.} 2023, Nature Astronomy

\bibitem[{{Dannert} {et~al.}(2022){Dannert}, {Ottiger}, {Quanz}, {Laugier}, {Fontanet}, {Gheorghe}, {Absil}, {Dandumont}, {Defr{\`e}re}, {Gasc{\'o}n}, {Glauser}, {Kammerer}, {Lichtenberg}, {Linz}, {Loicq}, \& {LIFE Collaboration}}]{dannert_et_al._2022}
{Dannert}, F.~A., {Ottiger}, M., {Quanz}, S.~P., {et~al.} 2022, \aap, 664, A22

\bibitem[{{Defr{\`e}re} {et~al.}(2010){Defr{\`e}re}, {Absil}, {den Hartog}, {Hanot}, \& {Stark}}]{2010A&A...509A...9D}
{Defr{\`e}re}, D., {Absil}, O., {den Hartog}, R., {Hanot}, C., \& {Stark}, C. 2010, \aap, 509, A9

\bibitem[{{Defr{\`e}re} {et~al.}(2021){Defr{\`e}re}, {Hinz}, {Kennedy}, {Stone}, {Rigley}, {Ertel}, {Gaspar}, {Bailey}, {Hoffmann}, {Mennesson}, {Millan-Gabet}, {Danchi}, {Absil}, {Arbo}, {Beichman}, {Bonavita}, {Brusa}, {Bryden}, {Downey}, {Esposito}, {Grenz}, {Haniff}, {Hill}, {Leisenring}, {Males}, {McMahon}, {Montoya}, {Morzinski}, {Pinna}, {Puglisi}, {Rieke}, {Roberge}, {Rousseau}, {Serabyn}, {Spalding}, {Skemer}, {Stapelfeldt}, {Su}, {Vaz}, {Weinberger}, \& {Wyatt}}]{2021AJ....161..186D}
{Defr{\`e}re}, D., {Hinz}, P.~M., {Kennedy}, G.~M., {et~al.} 2021, \aj, 161, 186

\bibitem[{{Dorn} \& {Lichtenberg}(2021)}]{2021ApJ...922L...4D}
{Dorn}, C. \& {Lichtenberg}, T. 2021, \apjl, 922, L4

\bibitem[{{Elkins-Tanton}(2012)}]{magma_oceans_2022}
{Elkins-Tanton}, L.~T. 2012, Annu Rev Earth Planet Sci, 40, 113

\bibitem[{Emsenhuber {et~al.}(2021)Emsenhuber, Mordasini, Burn, Alibert, Benz, \& Asphaug}]{Emsenhuber2021}
Emsenhuber, A., Mordasini, C., Burn, R., {et~al.} 2021, \aap, 656, A69

\bibitem[{{Ertel} {et~al.}(2018){Ertel}, {Defr{\`e}re}, {Hinz}, {Mennesson}, {Kennedy}, {Danchi}, {Gelino}, {Hill}, {Hoffmann}, {Rieke}, {Shannon}, {Spalding}, {Stone}, {Vaz}, {Weinberger}, {Willems}, {Absil}, {Arbo}, {Bailey}, {Beichman}, {Bryden}, {Downey}, {Durney}, {Esposito}, {Gaspar}, {Grenz}, {Haniff}, {Leisenring}, {Marion}, {McMahon}, {Millan-Gabet}, {Montoya}, {Morzinski}, {Pinna}, {Power}, {Puglisi}, {Roberge}, {Serabyn}, {Skemer}, {Stapelfeldt}, {Su}, {Vaitheeswaran}, \& {Wyatt}}]{2018AJ....155..194E}
{Ertel}, S., {Defr{\`e}re}, D., {Hinz}, P., {et~al.} 2018, \aj, 155, 194

\bibitem[{{Fortney} {et~al.}(2019){Fortney}, {Robinson}, {Domagal-Goldman}, {Genio}, {Gordon}, {Gharib-Nezhad}, {Lewis}, {Sousa-Silva}, {Airapetian}, {Drouin}, {Hargreaves}, {Huang}, {Karman}, {Ramirez}, {Rieker}, {Tennyson}, {Wordsworth}, {Yurchenko}, {Johnson}, {Lee}, {Marley}, {Dong}, {Kane}, {L{\'o}pez-Morales}, {Fauchez}, {Lee}, {Sung}, {Haghighipour}, {Horst}, {Gao}, {Kao}, {Dressing}, {Lupu}, {Savin}, {Fleury}, {Venot}, {Ascenzi}, {Milam}, {Linnartz}, {Gudipati}, {Gronoff}, {Salama}, {Gavilan}, {Bouwman}, {Turbet}, {Benilan}, {Henderson}, {Batalha}, {Jensen-Clem}, {Lyons}, {Freedman}, {Schwieterman}, {Goyal}, {Mancini}, {Irwin}, {Desert}, {Molaverdikhani}, {Gizis}, {Taylor}, {Lothringer}, {Pierrehumbert}, {Zellem}, {Batalha}, {Rugheimer}, {Lustig-Yaeger}, {Hu}, {Kempton}, {Arney}, {Line}, {Alam}, {Moses}, {Iro}, {Kreidberg}, {Blecic}, {Louden}, {Molli{\`e}re}, {Stevenson}, {Swain}, {Bott}, {Madhusudhan}, {Krissansen-Totton}, {Deming}, {Kitiashvili}, {Shkolnik}, {Rustamkulov}, {Rogers}, \&
  {Close}}]{2019astro2020T.146F}
{Fortney}, J., {Robinson}, T.~D., {Domagal-Goldman}, S., {et~al.} 2019, Astro2020, 2020, 146

\bibitem[{{Gaidos} {et~al.}(2019){Gaidos}, {Jacobs}, {LaCourse}, {Vanderburg}, {Rappaport}, {Berger}, {Pearce}, {Mann}, {Weiss}, {Fulton}, {Behmard}, {Howard}, {Ansdell}, {Ricker}, {Vanderspek}, {Latham}, {Seager}, {Winn}, \& {Jenkins}}]{2019MNRAS.488.4465G}
{Gaidos}, E., {Jacobs}, T., {LaCourse}, D., {et~al.} 2019, \mnras, 488, 4465

\bibitem[{{Gaidos} {et~al.}(2022){Gaidos}, {Mann}, {Rojas-Ayala}, {Feiden}, {Wood}, {Narayanan}, {Ansdell}, {Jacobs}, \& {LaCourse}}]{2022MNRAS.514.1386G}
{Gaidos}, E., {Mann}, A.~W., {Rojas-Ayala}, B., {et~al.} 2022, \mnras, 514, 1386

\bibitem[{{Gaillard} {et~al.}(2021){Gaillard}, {Bouhifd}, {F{\"u}ri}, {Malavergne}, {Marrocchi}, {Noack}, {Ortenzi}, {Roskosz}, \& {Vulpius}}]{2021SSRv..217...22G}
{Gaillard}, F., {Bouhifd}, M.~A., {F{\"u}ri}, E., {et~al.} 2021, \ssr, 217, 22

\bibitem[{{Goldblatt}(2015)}]{2015AsBio..15..362G}
{Goldblatt}, C. 2015, Astrobiology, 15, 362

\bibitem[{{Goldblatt} {et~al.}(2013){Goldblatt}, {Robinson}, {Zahnle}, \& {Crisp}}]{2013NatGe...6..661G}
{Goldblatt}, C., {Robinson}, T.~D., {Zahnle}, K.~J., \& {Crisp}, D. 2013, Nature Geoscience, 6, 661

\bibitem[{{Graham} {et~al.}(2021){Graham}, {Lichtenberg}, {Boukrouche}, \& {Pierrehumbert}}]{2021PSJ.....2..207G}
{Graham}, R.~J., {Lichtenberg}, T., {Boukrouche}, R., \& {Pierrehumbert}, R.~T. 2021, \psj, 2, 207

\bibitem[{{Hamano} {et~al.}(2013){Hamano}, {Abe}, \& {Genda}}]{2013Natur.497..607H}
{Hamano}, K., {Abe}, Y., \& {Genda}, H. 2013, \nat, 497, 607

\bibitem[{{Hamano} {et~al.}(2015){Hamano}, {Kawahara}, {Abe}, {Onishi}, \& {Hashimoto}}]{hamano_et_al._2015}
{Hamano}, K., {Kawahara}, H., {Abe}, Y., {Onishi}, M., \& {Hashimoto}, G.~L. 2015, \apj, 806, 216

\bibitem[{{Han} {et~al.}(2023){Han}, {Wyatt}, \& {Dent}}]{2023MNRAS.519.3257H}
{Han}, Y., {Wyatt}, M.~C., \& {Dent}, W.~R.~F. 2023, \mnras, 519, 3257

\bibitem[{Hansen \& Ireland(2022)}]{Hansen_2022}
Hansen, J.~T. \& Ireland, M.~J. 2022, \aap, 664, A52

\bibitem[{{Hansen} {et~al.}(2023){Hansen}, {Ireland}, {Laugier}, \& {LIFE Collaboration}}]{2023A&A...670A..57H}
{Hansen}, J.~T., {Ireland}, M.~J., {Laugier}, R., \& {LIFE Collaboration}. 2023, \aap, 670, A57

\bibitem[{{Hansen} {et~al.}(2022){Hansen}, {Ireland}, \& {LIFE Collaboration}}]{2022A&A...664A..52H}
{Hansen}, J.~T., {Ireland}, M.~J., \& {LIFE Collaboration}. 2022, \aap, 664, A52

\bibitem[{{Hirschmann}(2022)}]{2022GeCoA.328..221H}
{Hirschmann}, M.~M. 2022, \gca, 328, 221

\bibitem[{{Hirschmann}(2023)}]{2023E&PSL.61918311H}
{Hirschmann}, M.~M. 2023, Earth Planet Sci Lett, 619, 118311

\bibitem[{{Izidoro} {et~al.}(2021){Izidoro}, {Bitsch}, {Raymond}, {Johansen}, {Morbidelli}, {Lambrechts}, \& {Jacobson}}]{2021A&A...650A.152I}
{Izidoro}, A., {Bitsch}, B., {Raymond}, S.~N., {et~al.} 2021, \aap, 650, A152

\bibitem[{{Jackson} \& {Wyatt}(2012)}]{2012MNRAS.425..657J}
{Jackson}, A.~P. \& {Wyatt}, M.~C. 2012, \mnras, 425, 657

\bibitem[{{Janson} {et~al.}(2023){Janson}, {Patel}, {Ringqvist}, {Lu}, {Rebollido}, {Lichtenberg}, {Brandeker}, {Angerhausen}, \& {Noack}}]{2023A&A...671A.114J}
{Janson}, M., {Patel}, J., {Ringqvist}, S.~C., {et~al.} 2023, \aap, 671, A114

\bibitem[{{Janssen} {et~al.}(2023){Janssen}, {Woitke}, {Herbort}, {Min}, {Chubb}, {Helling}, \& {Carone}}]{2023AN....34430075J}
{Janssen}, L.~J., {Woitke}, P., {Herbort}, O., {et~al.} 2023, Astronomische Nachrichten, 344, e20230075

\bibitem[{{Johansen} {et~al.}(2021){Johansen}, {Ronnet}, {Bizzarro}, {Schiller}, {Lambrechts}, {Nordlund}, \& {Lammer}}]{2021SciA....7..444J}
{Johansen}, A., {Ronnet}, T., {Bizzarro}, M., {et~al.} 2021, Sci Adv, 7, eabc0444

\bibitem[{{Johansen} {et~al.}(2023){Johansen}, {Ronnet}, {Schiller}, {Deng}, \& {Bizzarro}}]{2023A&A...671A..75J}
{Johansen}, A., {Ronnet}, T., {Schiller}, M., {Deng}, Z., \& {Bizzarro}, M. 2023, \aap, 671, A75

\bibitem[{{Kammerer} \& {Quanz}(2018)}]{Kammerer_2018}
{Kammerer}, J. \& {Quanz}, S.~P. 2018, \aap, 609, A4

\bibitem[{{Kammerer} {et~al.}(2022){Kammerer}, {Quanz}, {Dannert}, \& {LIFE Collaboration}}]{2022A&A...668A..52K}
{Kammerer}, J., {Quanz}, S.~P., {Dannert}, F., \& {LIFE Collaboration}. 2022, \aap, 668, A52

\bibitem[{{Kane} {et~al.}(2019){Kane}, {Arney}, {Crisp}, {Domagal-Goldman}, {Glaze}, {Goldblatt}, {Grinspoon}, {Head}, {Lenardic}, {Unterborn}, {Way}, \& {Zahnle}}]{2019JGRE..124.2015K}
{Kane}, S.~R., {Arney}, G., {Crisp}, D., {et~al.} 2019, J Geophys Res Planets, 124, 2015

\bibitem[{{Kane} {et~al.}(2014){Kane}, {Kopparapu}, \& {Domagal-Goldman}}]{2014ApJ...794L...5K}
{Kane}, S.~R., {Kopparapu}, R.~K., \& {Domagal-Goldman}, S.~D. 2014, \apjl, 794, L5

\bibitem[{{Kegerreis} {et~al.}(2020){Kegerreis}, {Eke}, {Catling}, {Massey}, {Teodoro}, \& {Zahnle}}]{2020ApJ...901L..31K}
{Kegerreis}, J.~A., {Eke}, V.~R., {Catling}, D.~C., {et~al.} 2020, \apjl, 901, L31

\bibitem[{{Kegerreis} {et~al.}(2018){Kegerreis}, {Teodoro}, {Eke}, {Massey}, {Catling}, {Fryer}, {Korycansky}, {Warren}, \& {Zahnle}}]{2018ApJ...861...52K}
{Kegerreis}, J.~A., {Teodoro}, L.~F.~A., {Eke}, V.~R., {et~al.} 2018, \apj, 861, 52

\bibitem[{{Kempton} \& {Knutson}(2024)}]{Kempton2024}
{Kempton}, E. M.~R. \& {Knutson}, H.~A. 2024, Reviews in Mineralogy and Geochemistry, 90, 411

\bibitem[{{Kempton} {et~al.}(2023){Kempton}, {Lessard}, {Malik}, {Rogers}, {Futrowsky}, {Ih}, {Marounina}, \& {Mu{\~n}oz-Romero}}]{Kempton2023}
{Kempton}, E. M.~R., {Lessard}, M., {Malik}, M., {et~al.} 2023, \apj, 953, 57

\bibitem[{{Kennedy} {et~al.}(2015){Kennedy}, {Wyatt}, {Bailey}, {Bryden}, {Danchi}, {Defr{\`e}re}, {Haniff}, {Hinz}, {Lebreton}, {Mennesson}, {Millan-Gabet}, {Morales}, {Pani{\'c}}, {Rieke}, {Roberge}, {Serabyn}, {Shannon}, {Skemer}, {Stapelfeldt}, {Su}, \& {Weinberger}}]{kennedy2015}
{Kennedy}, G.~M., {Wyatt}, M.~C., {Bailey}, V., {et~al.} 2015, \apjs, 216, 23

\bibitem[{{Kenworthy} {et~al.}(2023){Kenworthy}, {Lock}, {Kennedy}, {van Capelleveen}, {Mamajek}, {Carone}, {Hambsch}, {Masiero}, {Mainzer}, {Kirkpatrick}, {Gomez}, {Leinhardt}, {Dou}, {Tanna}, {Sainio}, {Barker}, {Charbonnel}, {Garde}, {Le D{\^u}}, {Mulato}, {Petit}, \& {Rizzo Smith}}]{2023Natur.622..251K}
{Kenworthy}, M., {Lock}, S., {Kennedy}, G., {et~al.} 2023, \nat, 622, 251

\bibitem[{{Konrad} {et~al.}(2022){Konrad}, {Alei}, {Quanz}, {Angerhausen}, {Carri{\'o}n-Gonz{\'a}lez}, {Fortney}, {Grenfell}, {Kitzmann}, {Molli{\`e}re}, {Rugheimer}, {Wunderlich}, \& {LIFE Collaboration}}]{2022A&A...664A..23K}
{Konrad}, B.~S., {Alei}, E., {Quanz}, S.~P., {et~al.} 2022, \aap, 664, A23

\bibitem[{{Konrad} {et~al.}(2023){Konrad}, {Alei}, {Quanz}, {Molli{\`e}re}, {Angerhausen}, {Fortney}, {Hakim}, {Jordan}, {Kitzmann}, {Rugheimer}, {Shorttle}, {Wordsworth}, \& {LIFE Collaboration}}]{2023A&A...673A..94K}
{Konrad}, B.~S., {Alei}, E., {Quanz}, S.~P., {et~al.} 2023, \aap, 673, A94

\bibitem[{{Kral} {et~al.}(2017){Kral}, {Krivov}, {Defr{\`e}re}, {van Lieshout}, {Bonsor}, {Augereau}, {Th{\'e}bault}, {Ertel}, {Lebreton}, \& {Absil}}]{2017AstRv..13...69K}
{Kral}, Q., {Krivov}, A.~V., {Defr{\`e}re}, D., {et~al.} 2017, Astron Rev, 13, 69

\bibitem[{{Kral} {et~al.}(2020){Kral}, {Matr{\`a}}, {Kennedy}, {Marino}, \& {Wyatt}}]{2020MNRAS.497.2811K}
{Kral}, Q., {Matr{\`a}}, L., {Kennedy}, G.~M., {Marino}, S., \& {Wyatt}, M.~C. 2020, \mnras, 497, 2811

\bibitem[{{Kral} {et~al.}(2018){Kral}, {Wyatt}, {Triaud}, {Marino}, {Th{\'e}bault}, \& {Shorttle}}]{2018MNRAS.479.2649K}
{Kral}, Q., {Wyatt}, M.~C., {Triaud}, A. H.~M.~J., {et~al.} 2018, \mnras, 479, 2649

\bibitem[{{Krijt} {et~al.}(2023){Krijt}, {Kama}, {McClure}, {Teske}, {Bergin}, {Shorttle}, {Walsh}, \& {Raymond}}]{2023ASPC..534.1031K}
{Krijt}, S., {Kama}, M., {McClure}, M., {et~al.} 2023, in Astronomical Society of the Pacific Conference Series, Vol. 534, Protostars and Planets VII, ed. S.~{Inutsuka}, Y.~{Aikawa}, T.~{Muto}, K.~{Tomida}, \& M.~{Tamura}, 1031

\bibitem[{Lambrechts {et~al.}(2019)Lambrechts, Morbidelli, Jacobson, Johansen, Bitsch, Izidoro, \& Raymond}]{Lambrechts2019b}
Lambrechts, M., Morbidelli, A., Jacobson, S.~A., {et~al.} 2019, \aap, 627, A83

\bibitem[{{Lammer} {et~al.}(2020){Lammer}, {Scherf}, {Kurokawa}, {Ueno}, {Burger}, {Maindl}, {Johnstone}, {Leizinger}, {Benedikt}, {Fossati}, {Kislyakova}, {Marty}, {Avice}, {Fegley}, \& {Odert}}]{2020SSRv..216...74L}
{Lammer}, H., {Scherf}, M., {Kurokawa}, H., {et~al.} 2020, \ssr, 216, 74

\bibitem[{{Lay}(2004)}]{Lay04}
{Lay}, O.~P. 2004, \ao, 43, 6100

\bibitem[{{Lay}(2006)}]{Lay06}
{Lay}, O.~P. 2006, in SPIE, Vol. 6268, Advances in Stellar Interferometry, 62681A

\bibitem[{{Leconte} {et~al.}(2013){Leconte}, {Forget}, {Charnay}, {Wordsworth}, \& {Pottier}}]{2013Natur.504..268L}
{Leconte}, J., {Forget}, F., {Charnay}, B., {Wordsworth}, R., \& {Pottier}, A. 2013, \nat, 504, 268

\bibitem[{{Leconte} {et~al.}(2015){Leconte}, {Forget}, \& {Lammer}}]{2015ExA....40..449L}
{Leconte}, J., {Forget}, F., \& {Lammer}, H. 2015, Exp Astron, 40, 449

\bibitem[{{Lichtenberg}(2021)}]{2021ApJ...914L...4L}
{Lichtenberg}, T. 2021, \apjl, 914, L4

\bibitem[{{Lichtenberg} {et~al.}(2021){Lichtenberg}, {Bower}, {Hammond}, {Boukrouche}, {Sanan}, {Tsai}, \& {Pierrehumbert}}]{2021JGRE..12606711L}
{Lichtenberg}, T., {Bower}, D.~J., {Hammond}, M., {et~al.} 2021, J Geophys Res Planets, 126, e06711

\bibitem[{{Lichtenberg} \& {Clement}(2022)}]{2022ApJ...938L...3L}
{Lichtenberg}, T. \& {Clement}, M.~S. 2022, \apjl, 938, L3

\bibitem[{{Lichtenberg} \& {Miguel}(2024)}]{LichtenbergMiguel2024}
{Lichtenberg}, T. \& {Miguel}, Y. 2024, in Treatise on Geochemistry, 3rd ed., Treatise on Geochemistry, 3rd ed., arXiv:2405.04057

\bibitem[{{Lichtenberg} {et~al.}(2023){Lichtenberg}, {Schaefer}, {Nakajima}, \& {Fischer}}]{2023ASPC..534..907L}
{Lichtenberg}, T., {Schaefer}, L.~K., {Nakajima}, M., \& {Fischer}, R.~A. 2023, in Astronomical Society of the Pacific Conference Series, Vol. 534, Protostars and Planets VII, ed. S.~{Inutsuka}, Y.~{Aikawa}, T.~{Muto}, K.~{Tomida}, \& M.~{Tamura}, 907

\bibitem[{{Lock} {et~al.}(2020){Lock}, {Stewart}, \& {{\'C}uk}}]{2020E&PSL.53015885L}
{Lock}, S.~J., {Stewart}, S.~T., \& {{\'C}uk}, M. 2020, EPSL, 530, 115885

\bibitem[{{Lock} {et~al.}(2018){Lock}, {Stewart}, {Petaev}, {Leinhardt}, {Mace}, {Jacobsen}, \& {Cuk}}]{2018JGRE..123..910L}
{Lock}, S.~J., {Stewart}, S.~T., {Petaev}, M.~I., {et~al.} 2018, J Geophys Res Planets, 123, 910

\bibitem[{{L{\"o}hne} {et~al.}(2008){L{\"o}hne}, {Krivov}, \& {Rodmann}}]{2008ApJ...673.1123L}
{L{\"o}hne}, T., {Krivov}, A.~V., \& {Rodmann}, J. 2008, \apj, 673, 1123

\bibitem[{Luger \& Barnes(2015)}]{Luger2015}
Luger, R. \& Barnes, R. 2015, Astrobiology, 15, 119

\bibitem[{{Luger} \& {Barnes}(2015)}]{2015AsBio..15..119L}
{Luger}, R. \& {Barnes}, R. 2015, Astrobiology, 15, 119

\bibitem[{{Lupu} {et~al.}(2014){Lupu}, {Zahnle}, {Marley}, {Schaefer}, {Fegley}, {Morley}, {Cahoy}, {Freedman}, \& {Fortney}}]{2014ApJ...784...27L}
{Lupu}, R.~E., {Zahnle}, K., {Marley}, M.~S., {et~al.} 2014, \apj, 784, 27

\bibitem[{{Mamajek}(2009)}]{2009AIPC.1158....3M}
{Mamajek}, E.~E. 2009, in American Institute of Physics Conference Series, Vol. 1158, Exoplanets and Disks: Their Formation and Diversity, ed. T.~{Usuda}, M.~{Tamura}, \& M.~{Ishii} (AIP), 3--10

\bibitem[{{Mamajek}(2016)}]{Mamajek_2016}
{Mamajek}, E.~E. 2016, in Young Stars \& Planets Near the Sun, ed. J.~H. {Kastner}, B.~{Stelzer}, \& S.~A. {Metchev}, Vol. 314, 21--26

\bibitem[{{Mamajek} \& {Meyer}(2007)}]{2007ApJ...668L.175M}
{Mamajek}, E.~E. \& {Meyer}, M.~R. 2007, \apjl, 668, L175

\bibitem[{{Matsuo} {et~al.}(2023){Matsuo}, {Dannert}, {Laugier}, {Quanz}, {Kova{\v{c}}evi{\'c}}, \& {LIFE Collaboration}}]{2023A&A...678A..97M}
{Matsuo}, T., {Dannert}, F., {Laugier}, R., {et~al.} 2023, \aap, 678, A97

\bibitem[{{Meier} {et~al.}(2023){Meier}, {Bower}, {Lichtenberg}, {Hammond}, \& {Tackley}}]{2023A&A...678A..29M}
{Meier}, T.~G., {Bower}, D.~J., {Lichtenberg}, T., {Hammond}, M., \& {Tackley}, P.~J. 2023, \aap, 678, A29

\bibitem[{{Meier} {et~al.}(2021){Meier}, {Bower}, {Lichtenberg}, {Tackley}, \& {Demory}}]{2021ApJ...908L..48M}
{Meier}, T.~G., {Bower}, D.~J., {Lichtenberg}, T., {Tackley}, P.~J., \& {Demory}, B.-O. 2021, \apjl, 908, L48

\bibitem[{{Miller-Ricci} {et~al.}(2009){Miller-Ricci}, {Meyer}, {Seager}, \& {Elkins-Tanton}}]{2009ApJ...704..770M}
{Miller-Ricci}, E., {Meyer}, M.~R., {Seager}, S., \& {Elkins-Tanton}, L. 2009, \apj, 704, 770

\bibitem[{Morbidelli \& Nesvorny(2012)}]{Morbidelli2012a}
Morbidelli, A. \& Nesvorny, D. 2012, \aap, 546, 1

\bibitem[{{Olson} {et~al.}(2022){Olson}, {Sharp}, \& {Garai}}]{2022E&PSL.58717537O}
{Olson}, P., {Sharp}, Z., \& {Garai}, S. 2022, EPSL, 587, 117537

\bibitem[{{Olson} \& {Sharp}(2023)}]{2023E&PSL.62218418O}
{Olson}, P.~L. \& {Sharp}, Z.~D. 2023, EPSL, 622, 118418

\bibitem[{{Ostberg} {et~al.}(2023){Ostberg}, {Guzewich}, {Kane}, {Kohler}, {Oman}, {Fauchez}, {Kopparapu}, {Richardson}, \& {Whelley}}]{2023AJ....166..199O}
{Ostberg}, C.~M., {Guzewich}, S.~D., {Kane}, S.~R., {et~al.} 2023, \aj, 166, 199

\bibitem[{{Owen} \& {Mohanty}(2016)}]{2016MNRAS.459.4088O}
{Owen}, J.~E. \& {Mohanty}, S. 2016, \mnras, 459, 4088

\bibitem[{{Pierrehumbert} \& {Gaidos}(2011)}]{2011ApJ...734L..13P}
{Pierrehumbert}, R. \& {Gaidos}, E. 2011, \apjl, 734, L13

\bibitem[{{Piette} {et~al.}(2023){Piette}, {Gao}, {Brugman}, {Shahar}, {Lichtenberg}, {Miozzi}, \& {Driscoll}}]{2023ApJ...954...29P}
{Piette}, A. A.~A., {Gao}, P., {Brugman}, K., {et~al.} 2023, \apj, 954, 29

\bibitem[{{Pluriel} {et~al.}(2019){Pluriel}, {Marcq}, \& {Turbet}}]{2019Icar..317..583P}
{Pluriel}, W., {Marcq}, E., \& {Turbet}, M. 2019, \icarus, 317, 583

\bibitem[{{Quanz} {et~al.}(2022{\natexlab{a}}){Quanz}, {Absil}, {Benz}, {Bonfils}, {Berger}, {Defr{\`e}re}, {van Dishoeck}, {Ehrenreich}, {Fortney}, {Glauser}, {Grenfell}, {Janson}, {Kraus}, {Krause}, {Labadie}, {Lacour}, {Line}, {Linz}, {Loicq}, {Miguel}, {Pall{\'e}}, {Queloz}, {Rauer}, {Ribas}, {Rugheimer}, {Selsis}, {Snellen}, {Sozzetti}, {Stapelfeldt}, {Udry}, \& {Wyatt}}]{2022ExA....54.1197Q}
{Quanz}, S.~P., {Absil}, O., {Benz}, W., {et~al.} 2022{\natexlab{a}}, Exp Astron, 54, 1197

\bibitem[{{Quanz} {et~al.}(2022{\natexlab{b}}){Quanz}, {Ottiger}, {Fontanet}, {Kammerer}, {Menti}, {Dannert}, {Gheorghe}, {Absil}, {Airapetian}, {Alei}, {Allart}, {Angerhausen}, {Blumenthal}, {Buchhave}, {Cabrera}, {Carri{\'o}n-Gonz{\'a}lez}, {Chauvin}, {Danchi}, {Dandumont}, {Defr{\'e}re}, {Dorn}, {Ehrenreich}, {Ertel}, {Fridlund}, {Garc{\'\i}a Mu{\~n}oz}, {Gasc{\'o}n}, {Girard}, {Glauser}, {Grenfell}, {Guidi}, {Hagelberg}, {Helled}, {Ireland}, {Janson}, {Kopparapu}, {Korth}, {Kozakis}, {Kraus}, {L{\'e}ger}, {Leedj{\"a}rv}, {Lichtenberg}, {Lillo-Box}, {Linz}, {Liseau}, {Loicq}, {Mahendra}, {Malbet}, {Mathew}, {Mennesson}, {Meyer}, {Mishra}, {Molaverdikhani}, {Noack}, {Oza}, {Pall{\'e}}, {Parviainen}, {Quirrenbach}, {Rauer}, {Ribas}, {Rice}, {Romagnolo}, {Rugheimer}, {Schwieterman}, {Serabyn}, {Sharma}, {Stassun}, {Szul{\'a}gyi}, {Wang}, {Wunderlich}, {Wyatt}, \& {LIFE Collaboration}}]{quanz_et_al._2022}
{Quanz}, S.~P., {Ottiger}, M., {Fontanet}, E., {et~al.} 2022{\natexlab{b}}, \aap, 664, A21

\bibitem[{{Quintana} {et~al.}(2016){Quintana}, {Barclay}, {Borucki}, {Rowe}, \& {Chambers}}]{2016ApJ...821..126Q}
{Quintana}, E.~V., {Barclay}, T., {Borucki}, W.~J., {Rowe}, J.~F., \& {Chambers}, J.~E. 2016, \apj, 821, 126

\bibitem[{{Ramirez} \& {Kaltenegger}(2014)}]{2014ApJ...797L..25R}
{Ramirez}, R.~M. \& {Kaltenegger}, L. 2014, \apjl, 797, L25

\bibitem[{{Ramirez} \& {Kaltenegger}(2017)}]{2017ApJ...837L...4R}
{Ramirez}, R.~M. \& {Kaltenegger}, L. 2017, \apjl, 837, L4

\bibitem[{{Reiners} {et~al.}(2009){Reiners}, {Basri}, \& {Browning}}]{reiners_gibor_basri_matthew_2009}
{Reiners}, A., {Basri}, G., \& {Browning}, M. 2009, \apj, 692, 538

\bibitem[{{Rojas-Ayala} {et~al.}(2012){Rojas-Ayala}, {Covey}, {Muirhead}, \& {Lloyd}}]{bárbara_rojas-ayala_covey_muirhead_lloyd_2012}
{Rojas-Ayala}, B., {Covey}, K.~R., {Muirhead}, P.~S., \& {Lloyd}, J.~P. 2012, \apj, 748, 93

\bibitem[{{Salvador} {et~al.}(2017){Salvador}, {Massol}, {Davaille}, {Marcq}, {Sarda}, \& {Chassefi{\`e}re}}]{2017JGRE..122.1458S}
{Salvador}, A., {Massol}, H., {Davaille}, A., {et~al.} 2017, J Geophys Res Planets, 122, 1458

\bibitem[{{Salvador} \& {Samuel}(2023)}]{2023Icar..39015265S}
{Salvador}, A. \& {Samuel}, H. 2023, \icarus, 390, 115265

\bibitem[{{Sasselov} {et~al.}(2020){Sasselov}, {Grotzinger}, \& {Sutherland}}]{2020SciA....6.3419S}
{Sasselov}, D.~D., {Grotzinger}, J.~P., \& {Sutherland}, J.~D. 2020, Sci Adv, 6, eaax3419

\bibitem[{{Schaefer} {et~al.}(2016){Schaefer}, {Wordsworth}, {Berta-Thompson}, \& {Sasselov}}]{2016ApJ...829...63S}
{Schaefer}, L., {Wordsworth}, R.~D., {Berta-Thompson}, Z., \& {Sasselov}, D. 2016, \apj, 829, 63

\bibitem[{{Schlecker} {et~al.}(2024){Schlecker}, {Apai}, {Lichtenberg}, {Bergsten}, {Salvador}, \& {Hardegree-Ullman}}]{2024PSJ.....5....3S}
{Schlecker}, M., {Apai}, D., {Lichtenberg}, T., {et~al.} 2024, Planet Sci, 5, 3

\bibitem[{Schlecker {et~al.}(2021{\natexlab{a}})Schlecker, Mordasini, Emsenhuber, Klahr, Henning, Burn, Alibert, \& Benz}]{Schlecker2021}
Schlecker, M., Mordasini, C., Emsenhuber, A., {et~al.} 2021{\natexlab{a}}, \aap, 656, A71

\bibitem[{Schlecker {et~al.}(2021{\natexlab{b}})Schlecker, Pham, Burn, Alibert, Mordasini, Emsenhuber, Klahr, Henning, \& Mishra}]{Schlecker2021b}
Schlecker, M., Pham, D., Burn, R., {et~al.} 2021{\natexlab{b}}, \aap, 656, A73

\bibitem[{{Schlichting} \& {Young}(2022)}]{2022PSJ.....3..127S}
{Schlichting}, H.~E. \& {Young}, E.~D. 2022, \psj, 3, 127

\bibitem[{{Selsis} {et~al.}(2023){Selsis}, {Leconte}, {Turbet}, {Chaverot}, \& {Bolmont}}]{2023Natur.620..287S}
{Selsis}, F., {Leconte}, J., {Turbet}, M., {Chaverot}, G., \& {Bolmont}, {\'E}. 2023, \nat, 620, 287

\bibitem[{{Sleep} {et~al.}(2001){Sleep}, {Zahnle}, \& {Neuhoff}}]{2001PNAS...98.3666S}
{Sleep}, N.~H., {Zahnle}, K., \& {Neuhoff}, P.~S. 2001, PNAS, 98, 3666

\bibitem[{Solomatov(2015)}]{Solomatov_2015}
Solomatov, V. 2015, Treatise on Geophysics (Second Edition) (Eds. G. Schubert, Elsevier), 81

\bibitem[{{Stern}(1994)}]{1994AJ....108.2312S}
{Stern}, S.~A. 1994, \aj, 108, 2312

\bibitem[{{St{\"o}kl} {et~al.}(2015){St{\"o}kl}, {Dorfi}, \& {Lammer}}]{2015A&A...576A..87S}
{St{\"o}kl}, A., {Dorfi}, E., \& {Lammer}, H. 2015, \aap, 576, A87

\bibitem[{{Su} {et~al.}(2019){Su}, {Jackson}, {G{\'a}sp{\'a}r}, {Rieke}, {Dong}, {Olofsson}, {Kennedy}, {Leinhardt}, {Malhotra}, {Hammer}, {Meng}, {Rujopakarn}, {Rodriguez}, {Pepper}, {Reichart}, {James}, \& {Stassun}}]{2019AJ....157..202S}
{Su}, K. Y.~L., {Jackson}, A.~P., {G{\'a}sp{\'a}r}, A., {et~al.} 2019, \aj, 157, 202

\bibitem[{{Suer} {et~al.}(2023){Suer}, {Jackson}, {Grewal}, {Dalou}, \& {Lichtenberg}}]{2023FrEaS..1159412S}
{Suer}, T.-A., {Jackson}, C., {Grewal}, D.~S., {Dalou}, C., \& {Lichtenberg}, T. 2023, Front Earth Sci, 11, 1159412

\bibitem[{{Takai} {et~al.}(2008){Takai}, {Nakamura}, {Toki}, {Tsunogai}, {Miyazaki}, {Miyazaki}, {Hirayama}, {Nakagawa}, {Nunoura}, \& {Horikoshi}}]{2008PNAS..10510949T}
{Takai}, K., {Nakamura}, K., {Toki}, T., {et~al.} 2008, PNAS, 105, 10949

\bibitem[{{Thompson} {et~al.}(2019){Thompson}, {Weinberger}, {Keller}, {Arnold}, \& {Stark}}]{2019ApJ...875...45T}
{Thompson}, M.~A., {Weinberger}, A.~J., {Keller}, L.~D., {Arnold}, J.~A., \& {Stark}, C.~C. 2019, \apj, 875, 45

\bibitem[{{Turbet} {et~al.}(2021){Turbet}, {Bolmont}, {Chaverot}, {Ehrenreich}, {Leconte}, \& {Marcq}}]{2021Natur.598..276T}
{Turbet}, M., {Bolmont}, E., {Chaverot}, G., {et~al.} 2021, \nat, 598, 276

\bibitem[{{Turbet} {et~al.}(2019){Turbet}, {Ehrenreich}, {Lovis}, {Bolmont}, \& {Fauchez}}]{2019A&A...628A..12T}
{Turbet}, M., {Ehrenreich}, D., {Lovis}, C., {Bolmont}, E., \& {Fauchez}, T. 2019, \aap, 628, A12

\bibitem[{{Vican} \& {Schneider}(2014)}]{2014ApJ...780..154V}
{Vican}, L. \& {Schneider}, A. 2014, \apj, 780, 154

\bibitem[{{Way} \& {Del Genio}(2020)}]{2020JGRE..12506276W}
{Way}, M.~J. \& {Del Genio}, A.~D. 2020, J Geophys Res Planets, 125, e06276

\bibitem[{{Williams} \& {Mukhopadhyay}(2019)}]{2019Natur.565...78W}
{Williams}, C.~D. \& {Mukhopadhyay}, S. 2019, \nat, 565, 78

\bibitem[{{Wordsworth} \& {Kreidberg}(2022)}]{2022ARA&A..60..159W}
{Wordsworth}, R. \& {Kreidberg}, L. 2022, \araa, 60, 159

\bibitem[{{Wordsworth} {et~al.}(2018){Wordsworth}, {Schaefer}, \& {Fischer}}]{2018AJ....155..195W}
{Wordsworth}, R.~D., {Schaefer}, L.~K., \& {Fischer}, R.~A. 2018, \aj, 155, 195

\bibitem[{Wyatt {et~al.}(2019)Wyatt, Kral, \& Sinclair}]{Wyatt2019}
Wyatt, M.~C., Kral, Q., \& Sinclair, C.~A. 2019, Mon Not R Astron Soc, 802, 782

\bibitem[{{Wyatt} {et~al.}(2007){Wyatt}, {Smith}, {Greaves}, {Beichman}, {Bryden}, \& {Lisse}}]{2007ApJ...658..569W}
{Wyatt}, M.~C., {Smith}, R., {Greaves}, J.~S., {et~al.} 2007, \apj, 658, 569

\bibitem[{{Zahnle} \& {Carlson}(2020)}]{2020plas.book....3Z}
{Zahnle}, K.~J. \& {Carlson}, R.~W. 2020, in Planetary Astrobiology, ed. V.~S. {Meadows}, G.~N. {Arney}, B.~E. {Schmidt}, \& D.~J. {Des Marais} (University of Arizona Press), 3--36

\bibitem[{{Zhang} \& {Sigurdsson}(2003)}]{2003ApJ...596L..95Z}
{Zhang}, B. \& {Sigurdsson}, S. 2003, \apjl, 596, L95

\bibitem[{{Zhu} \& {Dong}(2021)}]{exostats}
{Zhu}, W. \& {Dong}, S. 2021, \araa, 59, 291

\end{thebibliography}

\rule{0.4\textwidth}{0.6pt}

\begin{appendix}
\section{Exozodi Dust Limits}
\label{sec: appendixA}
\hypertarget{appendix1}{}

We performed further simulations to study the effect of exozodiacal dust on LIFE's instrument to a greater extent. These plots are complementary to the ones presented in Sec. \ref{subsec: exozodi} to present a more complete study on the matter. 

Figure \ref{fig: A1} shows the resulting S/N contour plots for shorter integration times in the scenario from Fig. \ref{fig: 12} (c), namely a 1500 K planet orbiting a sun-like star at 1AU. While Fig. \ref{fig: A4} shows shorter integration times for the planet orbiting an M-dwarf at 0.13, from Fig. \ref{fig: 12} (d). Figures \ref{fig: A2}-\ref{fig: A3} show the resulting S/N contour plots for the situations displayed respectively in Fig. \ref{fig: 12} (a) and (b), with longer integration times. Namely a 700 K planet orbiting both a sun-like star at 1AU in Fig. \ref{fig: A2} and an M-dwarf at 0.13AU in Fig. \ref{fig: A3}. The integration times considered are 30, 50, 70 and 100 hours.

\subsection{G star}

\begin{figure}[h] 
\centering
\begin{center}
\subfloat[][]{\includegraphics[width=0.56\textwidth]{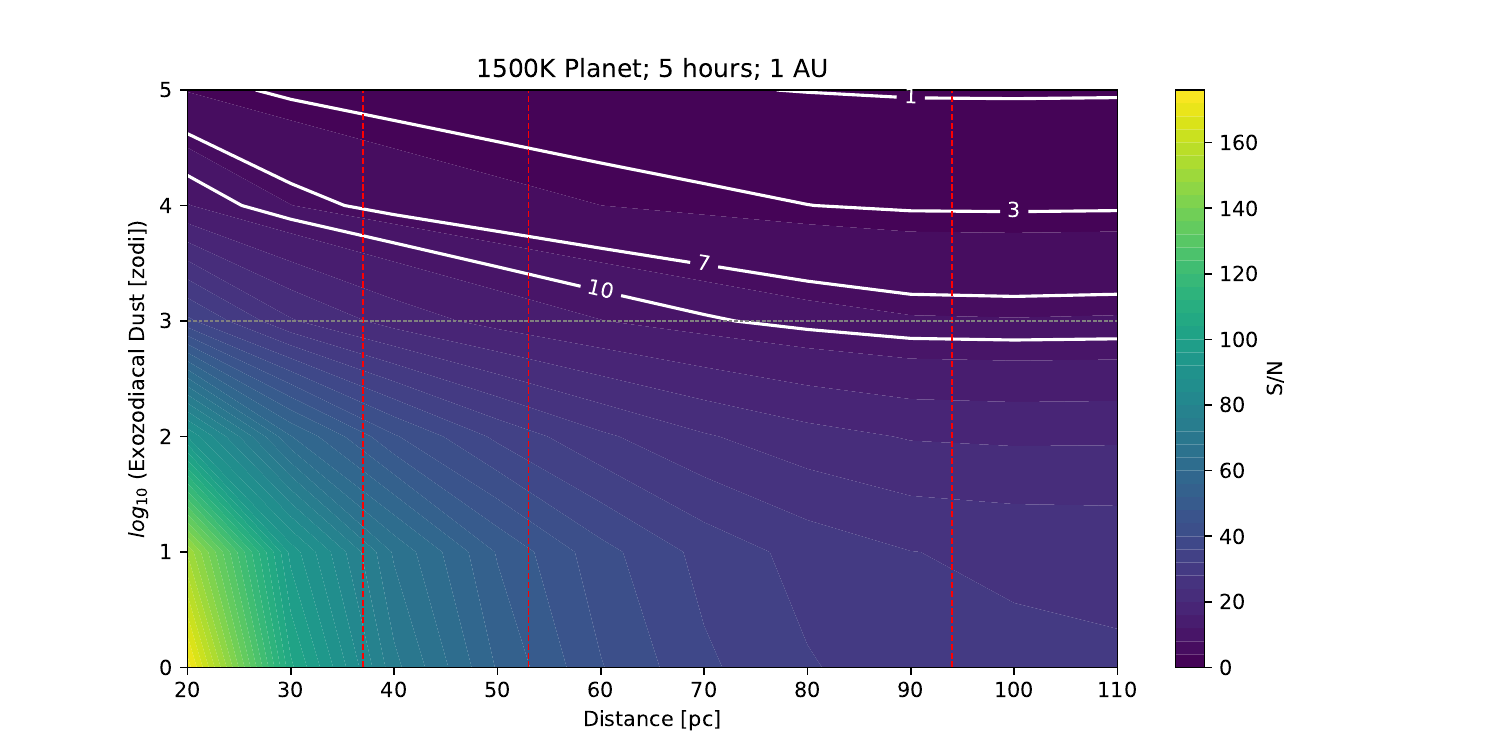}}\quad
\subfloat[][]{\includegraphics[width=0.56\textwidth]{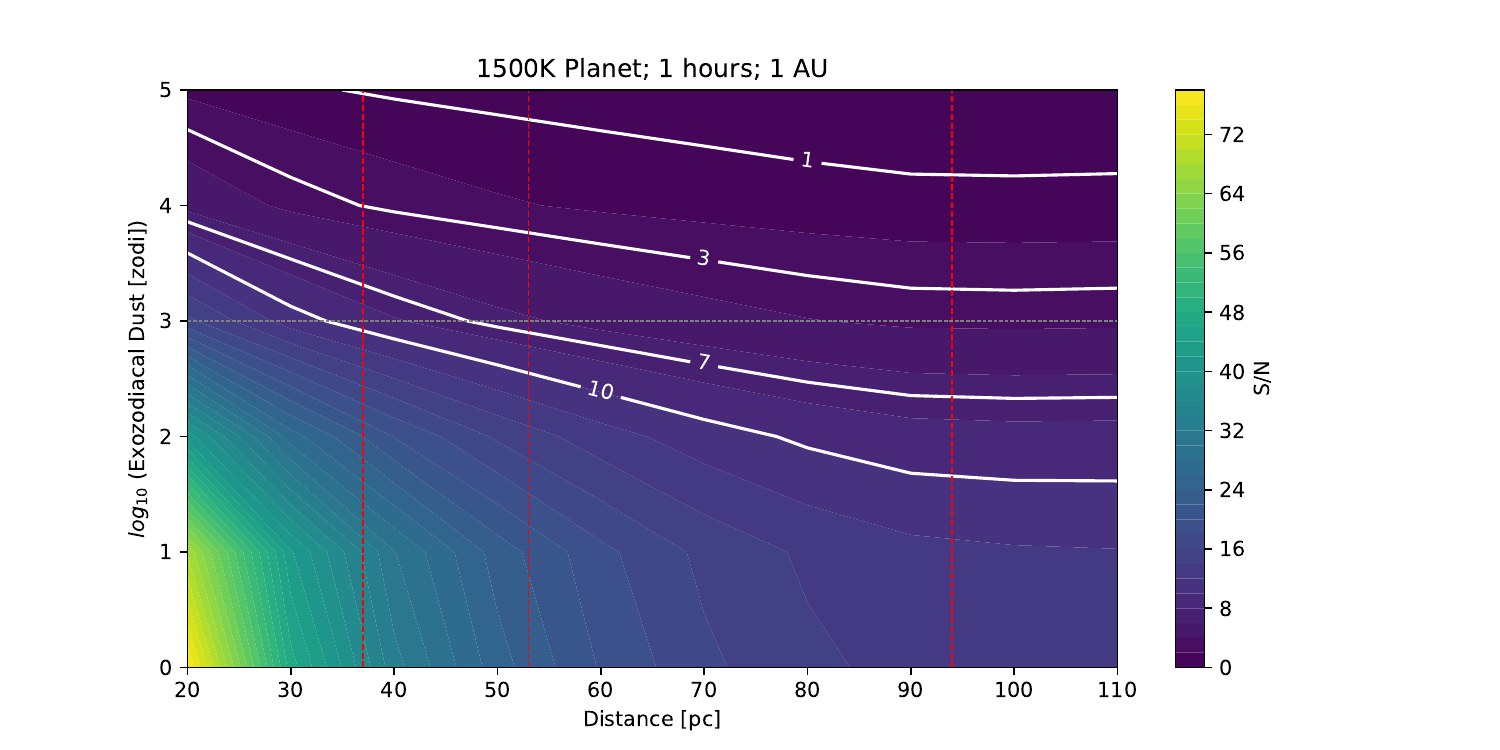}}\quad
\caption{Contour plots of S/N for a 1500 K planet orbiting a sun-like star at 1AU for (a) 5 and (b) 1 hours of integration time}
\label{fig: A1}
\end{center}
\end{figure}

\begin{figure}[h] 
\centering
\begin{center}
\subfloat[][]{\includegraphics[width=0.6\textwidth]{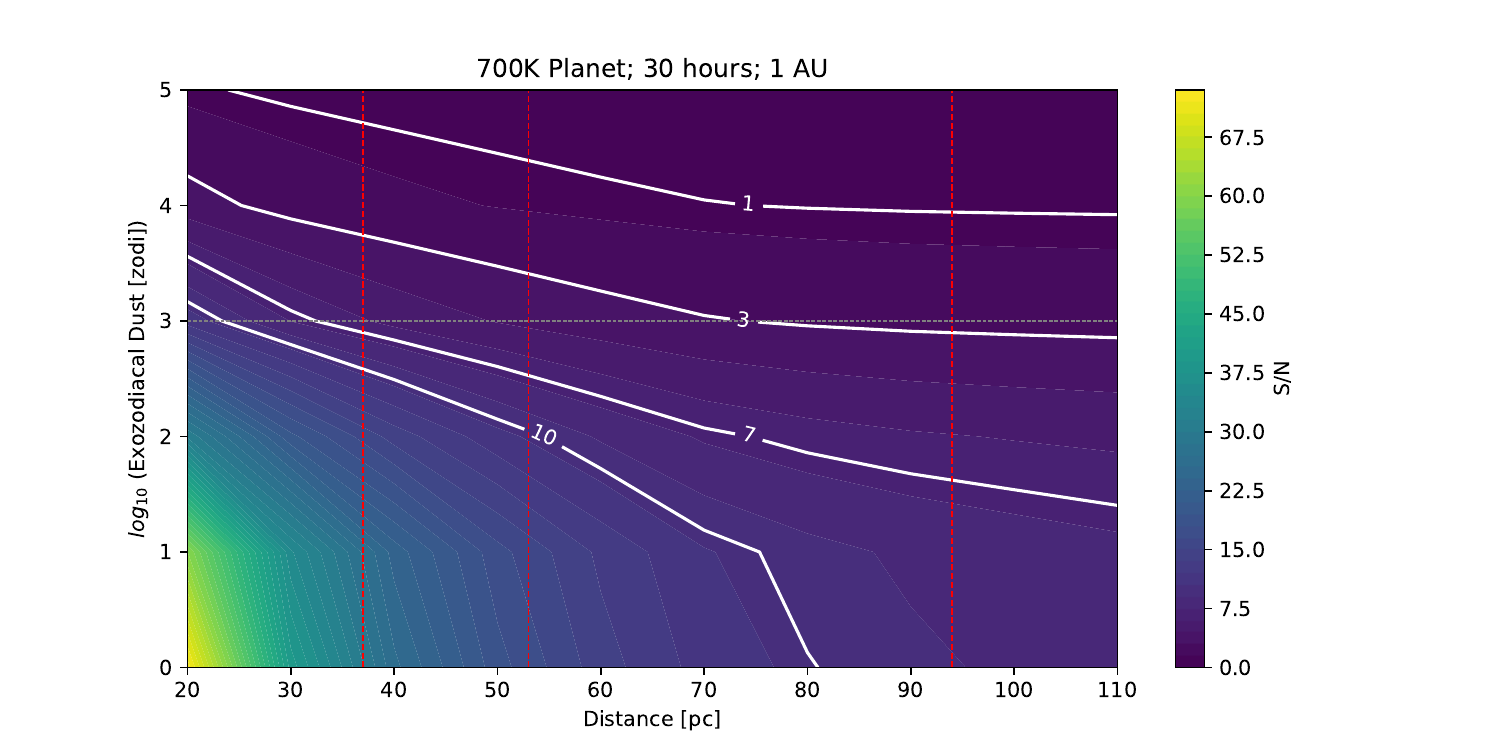}}\quad
\subfloat[][]{\includegraphics[width=0.6\textwidth]{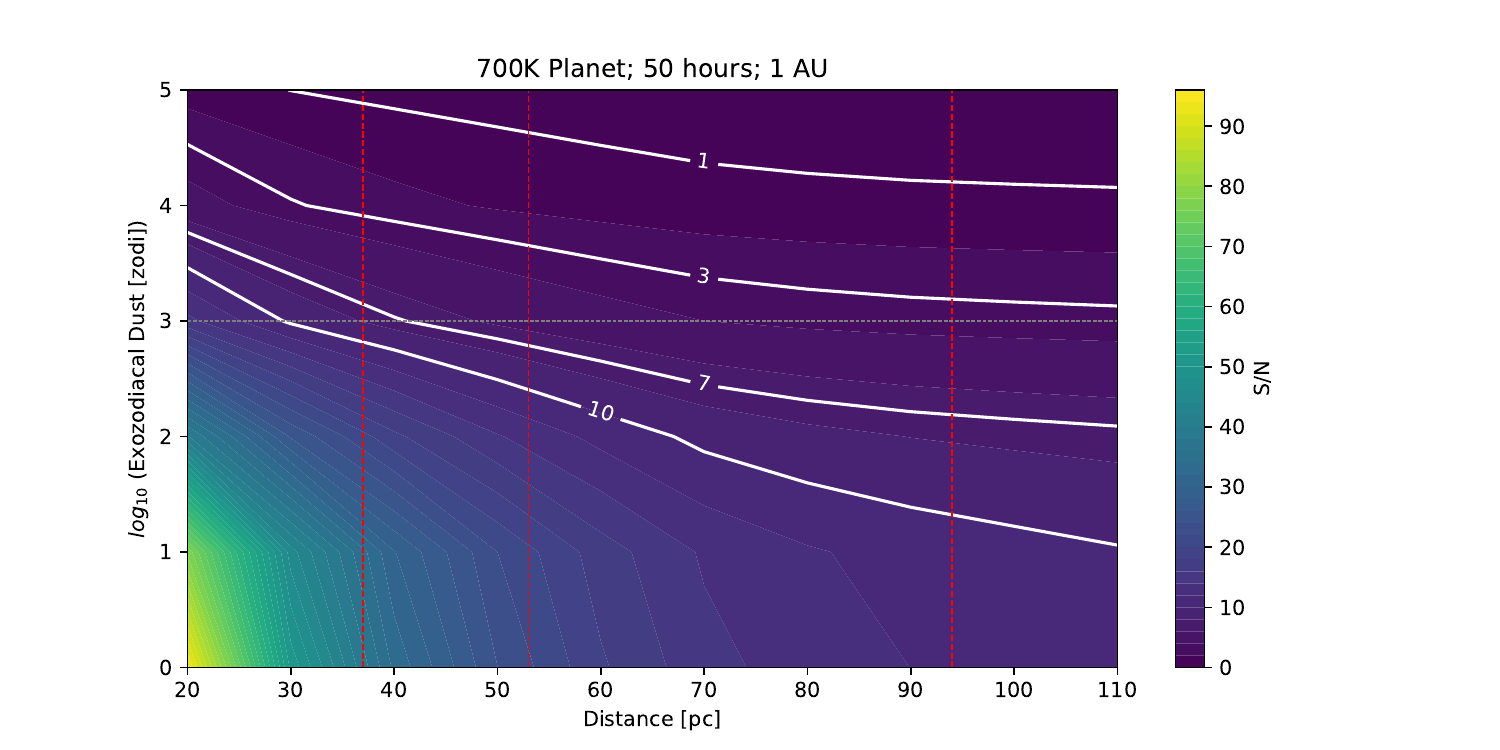}}\quad
\subfloat[][]{\includegraphics[width=0.6\textwidth]{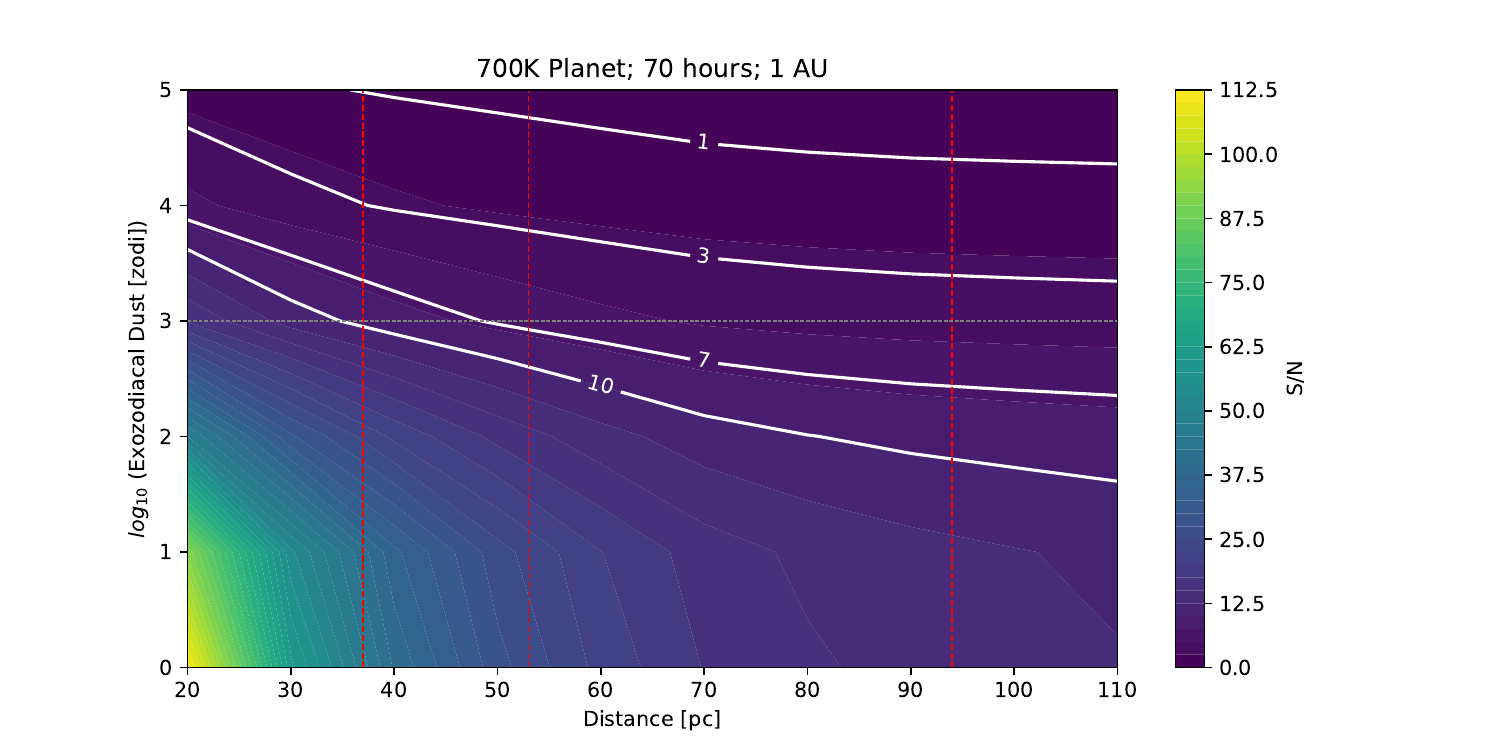}}\quad
\subfloat[][]{\includegraphics[width=0.6\textwidth]{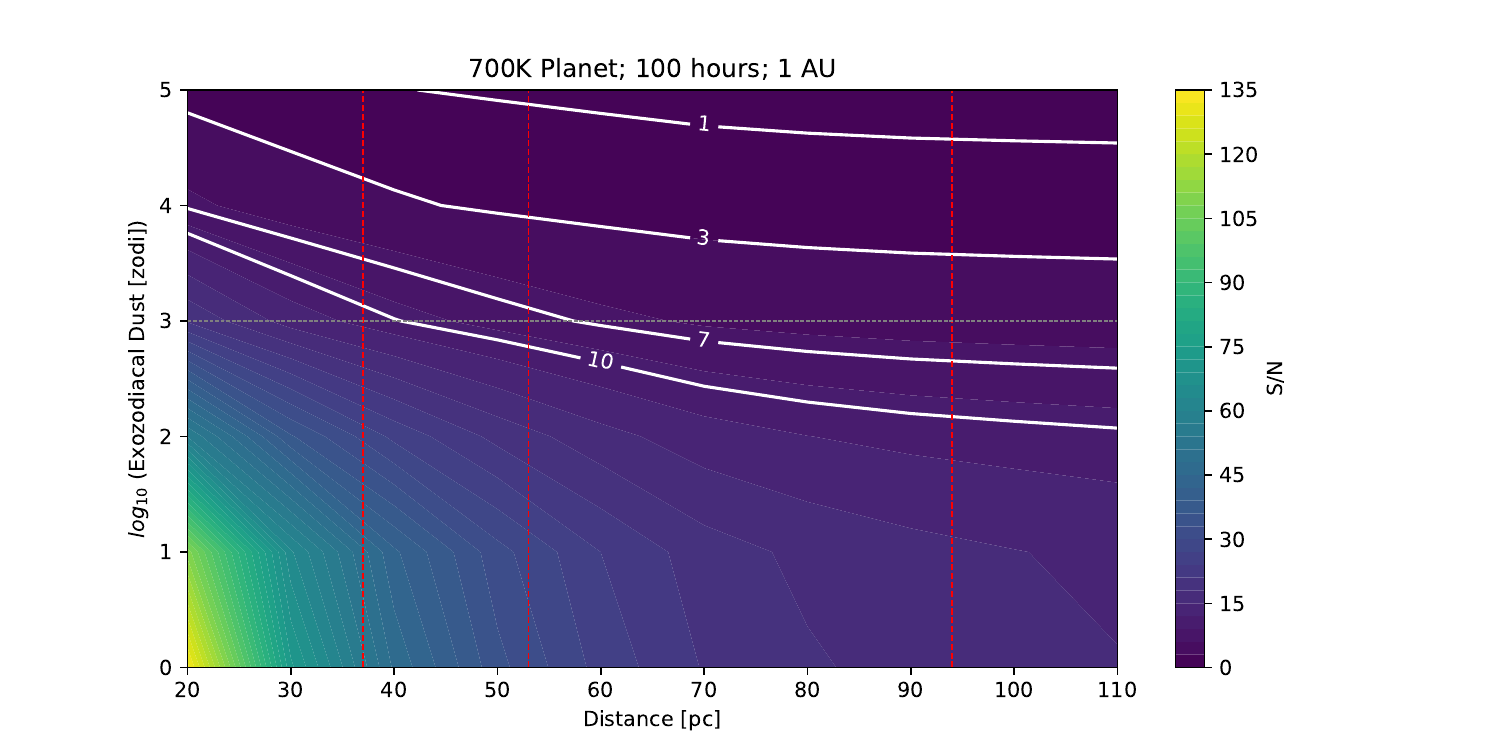}}\quad
\caption{Contour plots of S/N for a 700 K planet orbiting a sun-like star at 1AU for (a) 30, (b) 50, (c) 70, and (d) 100 hours of integration time}
\label{fig: A2}
\end{center}
\end{figure}

For a planet orbiting a sun-like star detection at the farthest association remains a possibility at thousands of zodis for 5 hours of observation time, as per Fig. \ref{fig: A1} (a). For 1 hour, Fig. \ref{fig: A1} (b), only $\beta$ Pictoris remains detectable at a thousand zodis. Nonetheless TW Hydrae is still detectable for levels of exozodiacal dust of several hundreds. From Fig. \ref{fig: A2} it becomes clear that exozodiacal dust poses a challenge for protoplanets with moderate emission temperatures. $\beta$ Pictoris becomes detectable at $10^3$ zodis after 50 hours, but TW Hydrae is only detected after 100 hours of integration time.

\clearpage
\subsection*{M-dwarf}

\begin{figure}[h] 
\centering
\begin{center}
\subfloat[][]{\includegraphics[width=0.6\textwidth]{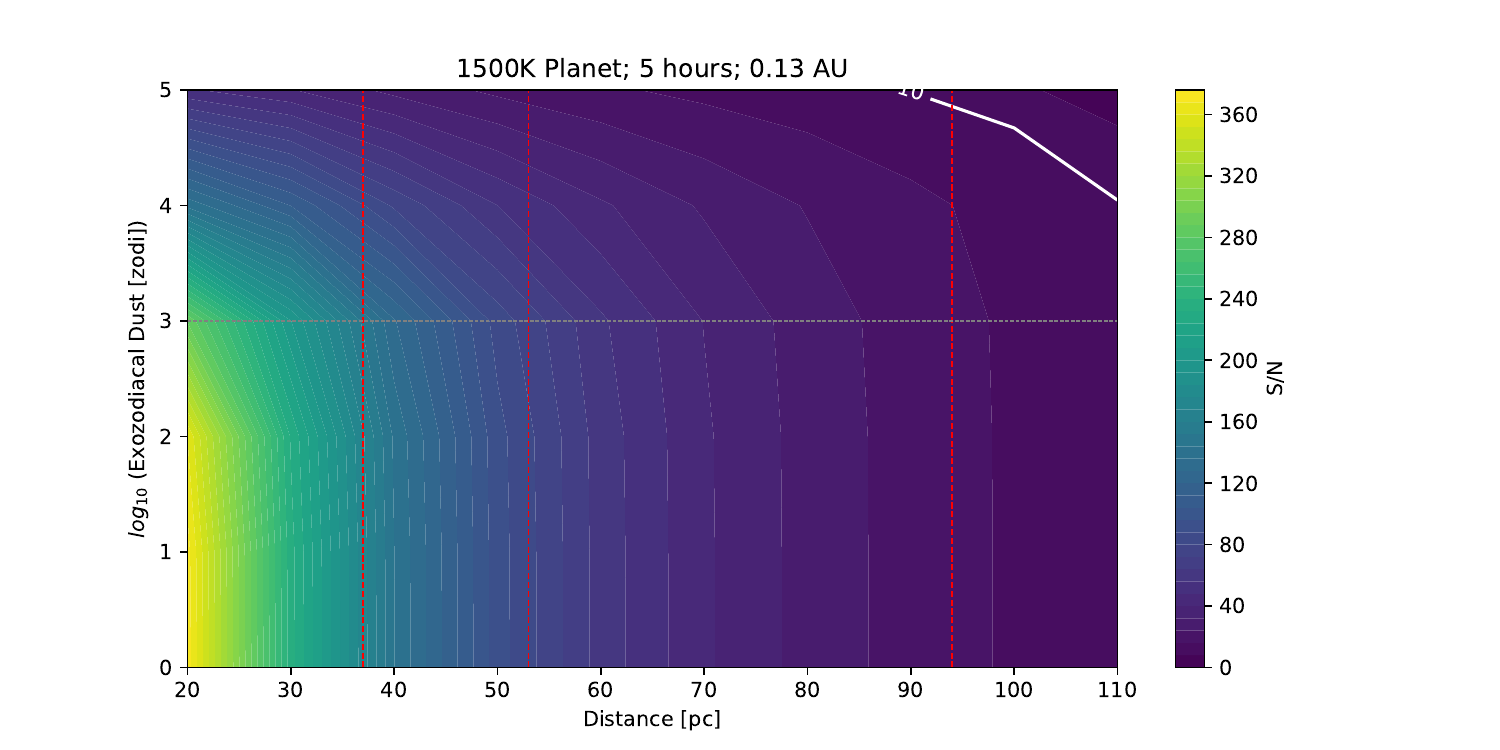}}\quad
\subfloat[][]{\includegraphics[width=0.6\textwidth]{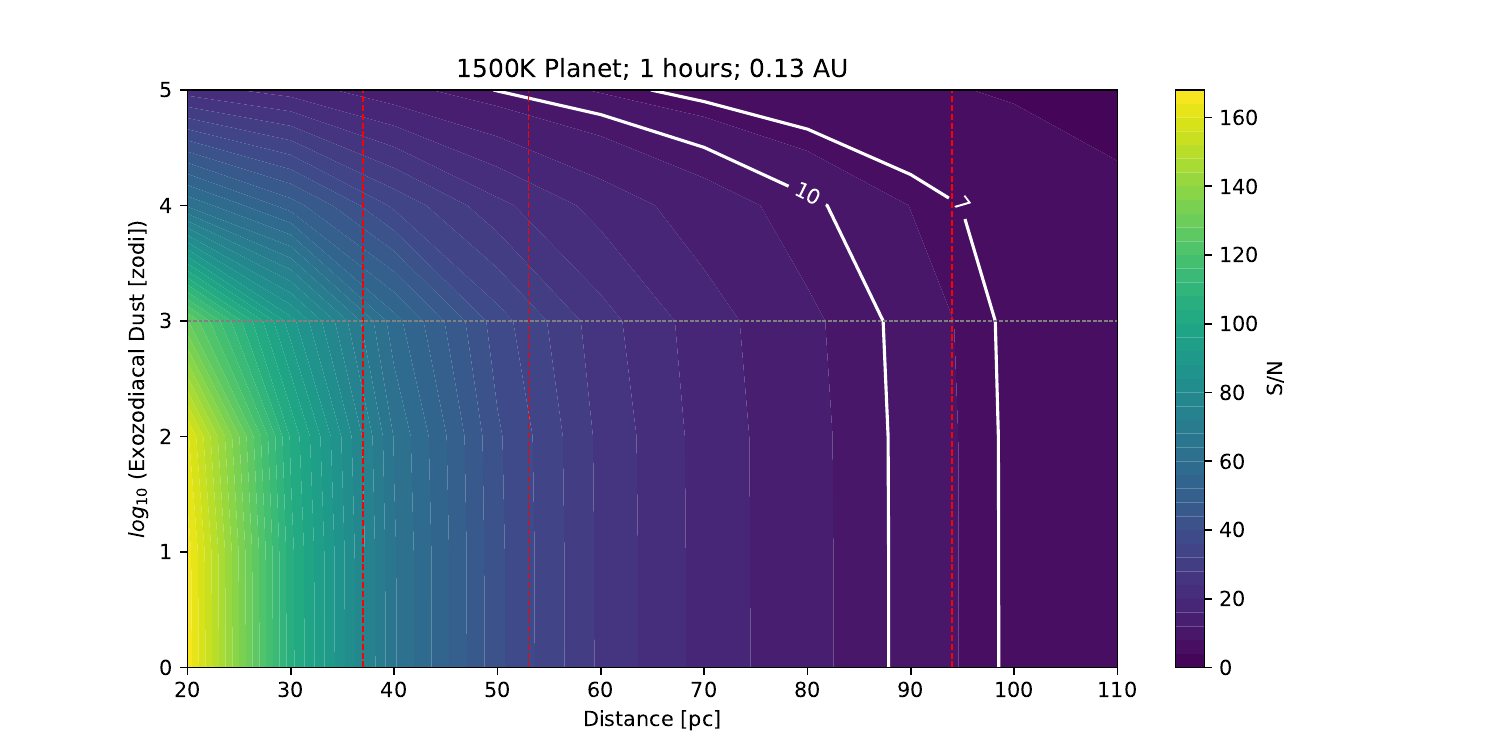}}\quad
\caption{Contour plots of S/N for a 1500 K planet orbiting an M-dwarf star at 0.13AU for (a) 5 and (b) 1 hours of integration time}
\label{fig: A3}
\end{center}
\end{figure}

\begin{figure}[h] 
\centering
\begin{center}
\subfloat[][]{\includegraphics[width=0.6\textwidth]{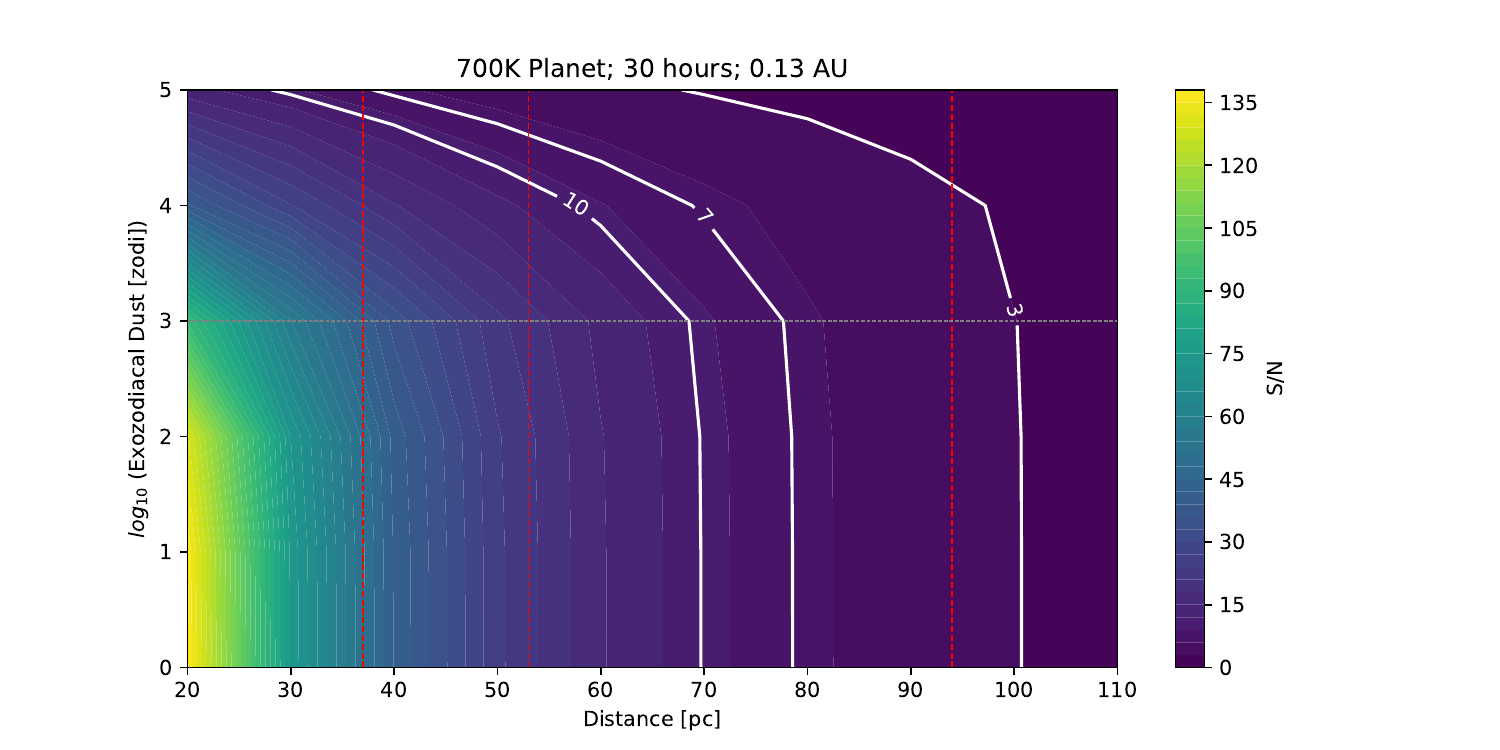}}\quad
\subfloat[][]{\includegraphics[width=0.6\textwidth]{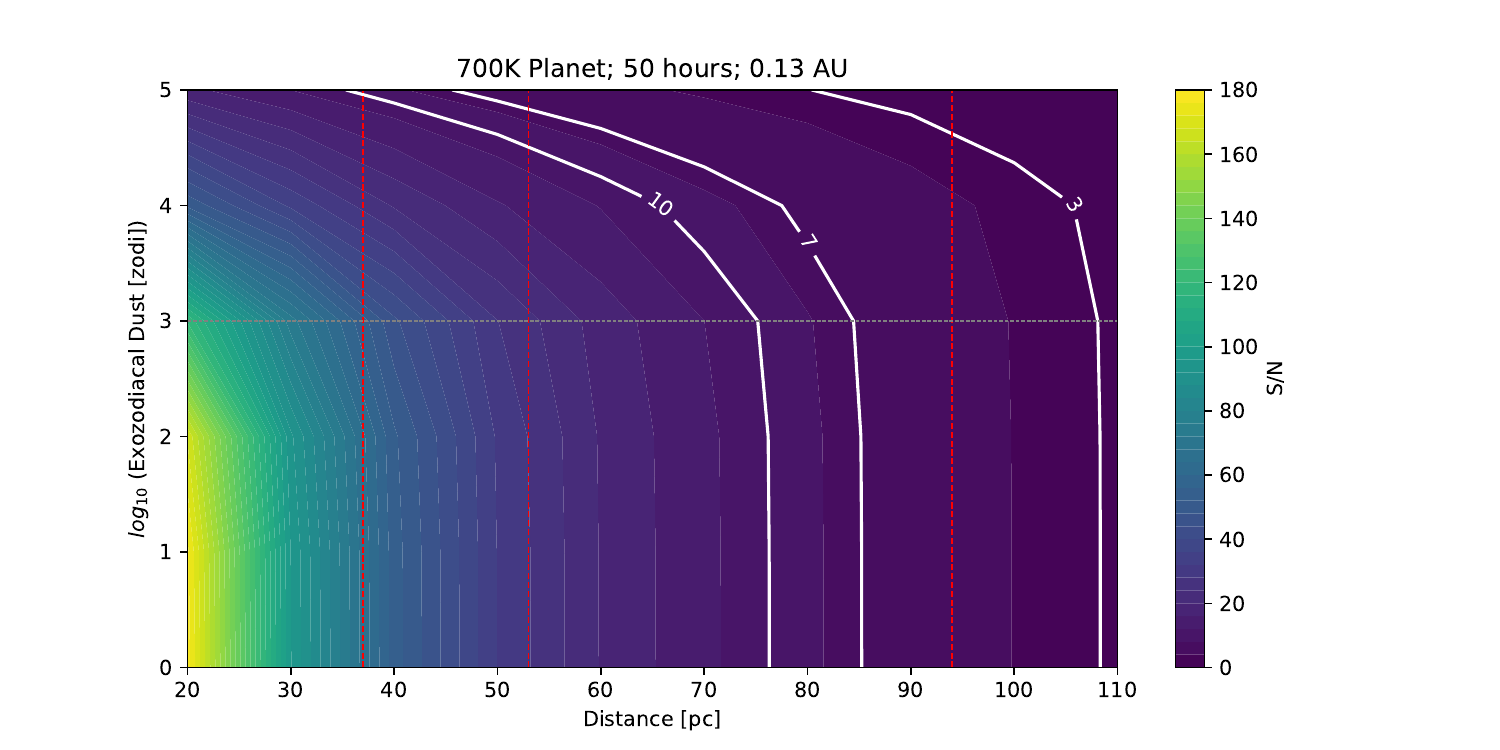}}\quad
\subfloat[][]{\includegraphics[width=0.6\textwidth]{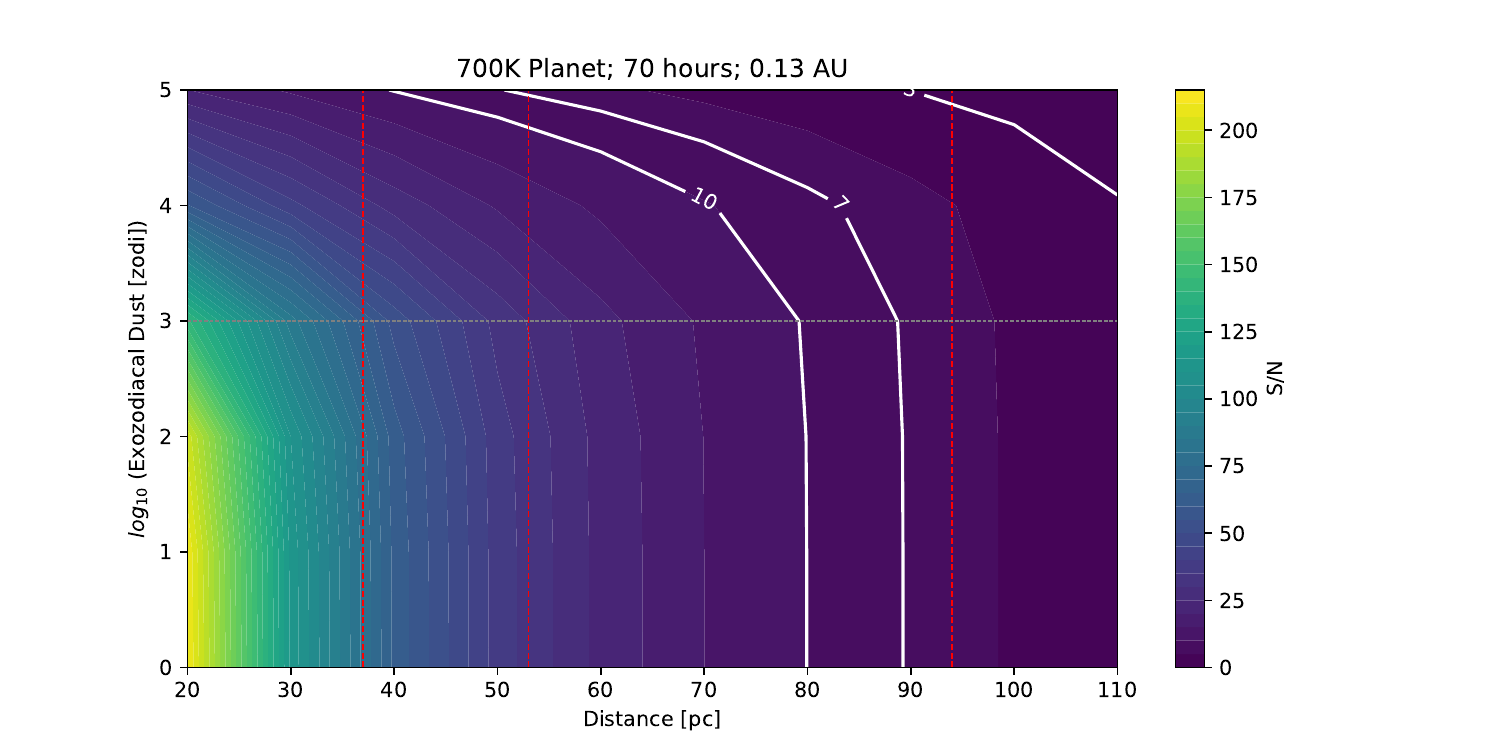}}\quad
\subfloat[][]{\includegraphics[width=0.6\textwidth]{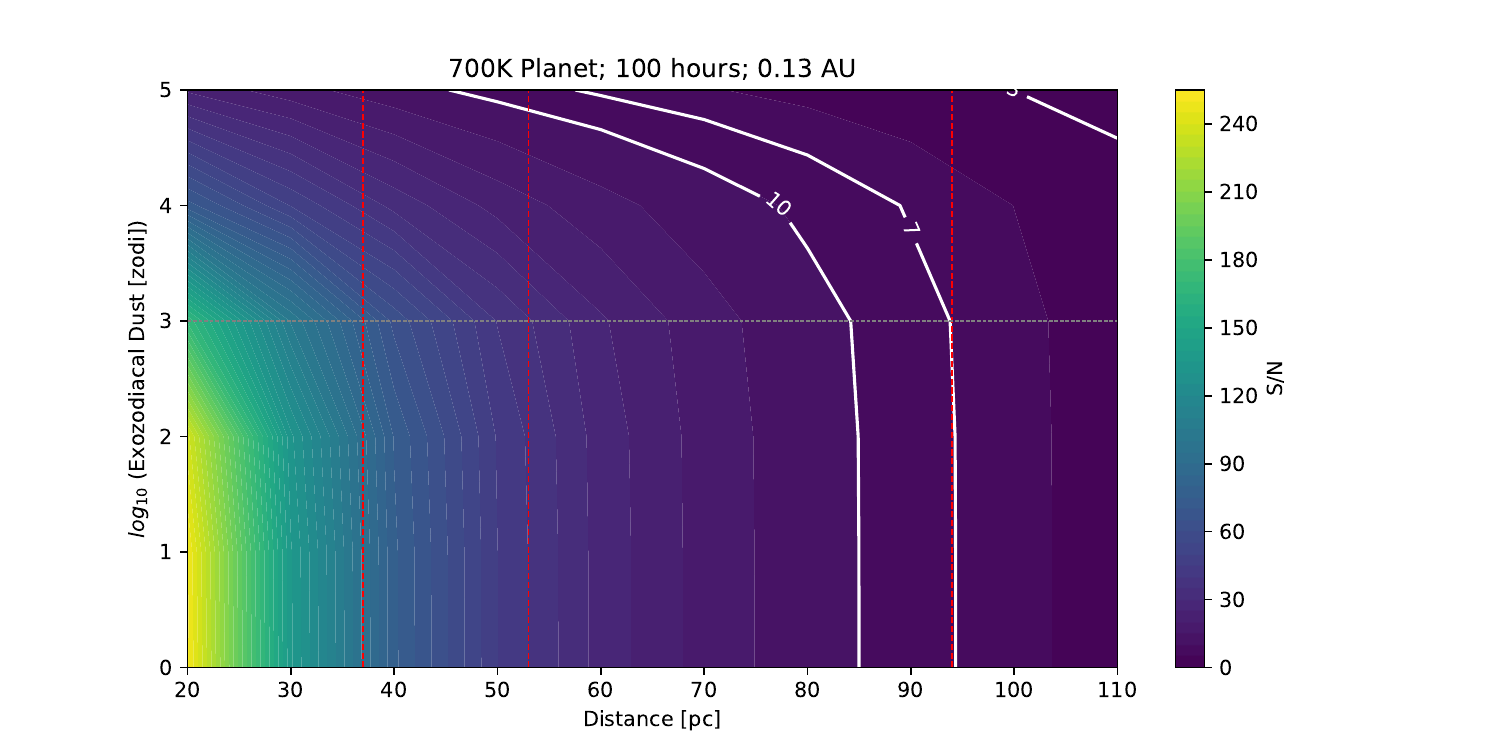}}\quad
\caption{Contour plots of S/N for a 700 K planet orbiting an M-dwarf star at 0.13AU for (a) 30, (b) 50, (c) 70, and (d) 100 hours of integration time}
\label{fig: A4}
\end{center}
\end{figure}

The situation for a planet orbiting an M-dwarf star at 0.13 is considerably better than its G-star counterpart. Figure \ref{fig: A3} shows that down to 1 hour of integration time, a planet in the farthest association $\eta$ Chamaeleontis is detectable for tens of thousands of zodis. Results from Fig. \ref{fig: A4} show that both $\beta$ Pictoris and TW Hydrae lie in the detectable region consistently for tens of thousands of zodis, as they were already with 10 hours of integration time. A planet in $\eta$ Chamaeleontis becomes detectable for a thousand zodis after 100 hours of integration time.

\clearpage

\section{Optimistic Throughput} \label{app:higherthroughput}

Throughout the simulations in this paper we kept the photon throughput parameter of the simulator fixed at $5\%$ (see Table \ref{tab:instrument_values}). This is a relatively pessimistic estimation for the instrument. Therefore it is worth to explore the instrument's performance in constraining the radiating temperature of the observed planet with an optimistic photon throughput of $20\%$. The following figures are reproductions of Figs. \ref{fig: 8}-\ref{fig: 10} respectively. The only difference is the throughput value in the set up of the instrument.

\begin{figure}[h] 
\centering
\begin{center}
\includegraphics[width=0.4\textwidth]{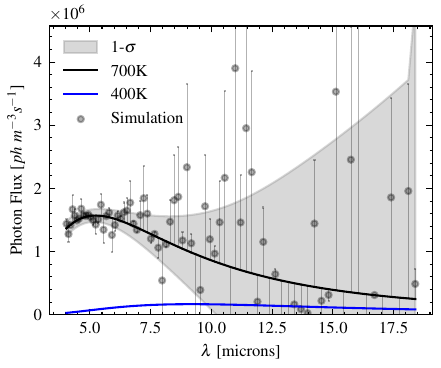}
\caption{Reproduction of Fig. \ref{fig: 8} with optimistic $20\%$ photon throughput.}
\label{fig: B1}
\end{center}
\end{figure}

\begin{figure}[h] 
\centering
\begin{center}
\includegraphics[width=0.4\textwidth]{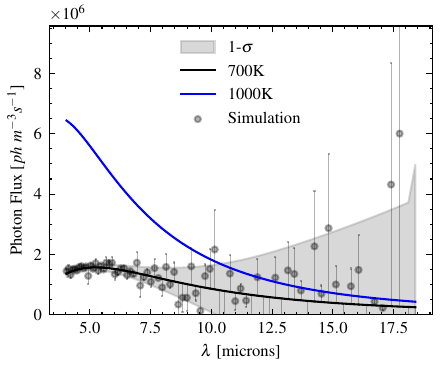}
\caption{Reproduction of Fig. \ref{fig: 9} with optimistic $20\%$ photon throughput.}
\label{fig: B2}
\end{center}
\end{figure}

\begin{figure}[h] 
\centering
\begin{center}
\includegraphics[width=0.4\textwidth]{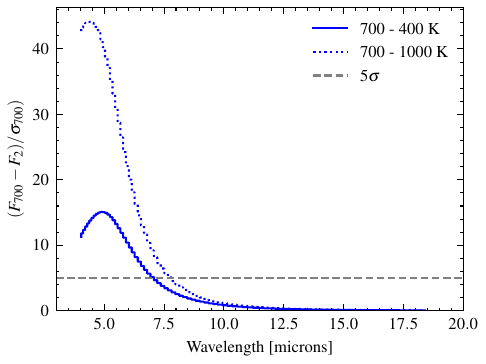}
\caption{Reproduction of Fig. \ref{fig: 10} with optimistic $20\%$ photon throughput.}
\label{fig: B3}
\end{center}
\end{figure}

The chances with a higher throughput are considerable more optimistic than in Sect. \ref{subsec: constraining}. In both Fig. \ref{fig: B1} and Fig. \ref{fig: B2} the photon flux of the planet being compared (blue line) reaches the 1-$\sigma$ uncertainty (shaded) area  at longer wavelengths than in their corresponding figures (Figs. \ref{fig: 8}-\ref{fig: 9}). In the paper we have already highlighted the importance of shorter wavelengths when observing magma ocean protoplanets with high temperatures ($\gtrsim$1000K). While longer wavelengths still contain a lot of uncertainty (wide shaded area), the uncertainty is restricted in the shorter wavelength regions. These results show how a greater photon throughput allows for a greater statistical difference between the considered fluxes (see Fig. \ref{fig: B3}) at shorter wavelengths. The peak of the curves in Fig. \ref{fig: B3} reach triple the sigma values for the brighter planet case (1000-700K) than in Fig. \ref{fig: 10} and double for the dimmer (400-700K) case. Furthermore, both curves pass the 5$\sigma$ threshold at longer wavelengths. This means that increasing the photon throughput increases the range of wavelengths useful for constraining radiating temperatures of observed planets.\\

Similar to these situations, we anticipate all other results presented in the paper to display analogous improvements in detection capability with a more optimistic throughput.
\end{appendix}
\end{document}